\documentclass{aa}  

\usepackage{graphicx}
\usepackage{txfonts}
\usepackage{xcolor}
\usepackage{ulem}
\usepackage{multicol}
\begin{document}

   \title{Long-term stellar activity of M dwarfs: \newline A combined \textit{K2} and TESS study of two early M-type stars}

   \author{St. Raetz 
          \and
          B. Stelzer
          }

   \institute{Institut f\"{u}r Astronomie und Astrophysik T\"{u}bingen, Eberhard Karls Universit\"{a}t T\"{u}bingen, Sand 1, D-72076 T\"{u}bingen, Germany\\
              \email{raetz@astro.uni-tuebingen.de}
         }

   \date{Received September 15, 1996; accepted March 16, 1997}

 \abstract{Studies of the rotation and activity of M type stars are essential to enhance our understanding of stellar dynamos and angular momentum evolution. Using the outstanding photometric capabilities of space telescopes rotation signals even with low amplitudes can be investigated in up to now unrivaled detail. By combining data of \textit{K2} and the TESS prime mission the star spot activity of M dwarfs can be monitored on half a decade timescale. In the framework of our study on the rotation-activity relation for bright and nearby M dwarfs we also aim at an investigation of the long-term activity. While \textit{K2} was observing fields distributed around the ecliptic plane, the TESS prime mission was oriented along a line of ecliptic longitude with one camera centered on an ecliptic pole. Due to these different observing strategies, the overlap between \textit{K2} and the TESS prime mission is marginal. However, 45 stars from our sample were observed with both missions of which two early M-type stars that fulfill our selection criteria, EPIC\,202059229 and EPIC\,245919787,  were analyzed in more detail. We found that for both stars the rotation period did not change while the rotational phase did change for EPIC\,245919787 by $\sim$0.2. The amplitude of the spot induced variability changed for both stars but more significant for EPIC\,245919787. By comparing the cumulative flare frequency distributions we found that the flare activity for EPIC\,202059229 is unchanged while it slightly changes for EPIC\,245919787 between the \textit{K2} and TESS epochs. Using a combination of light curves from \textit{K2} and TESS that span a baseline up to 4.5 years we could measure significant differential rotation for EPIC\,245919787. Furthermore, we show that combining missions like K2 and TESS is a promising method for detecting stellar activity cycles.}
 


   \maketitle
%

\section{Introduction}

Stellar activity is a direct consequence of strong magnetic fields affecting all layers of the stellar atmosphere. The observation of the phenomena of stellar activity e.g. stellar spots, flares as well long-term variability including activity cycles, is of crucial importance for advancing our understanding of active stars. Stellar activity is believed to be generated and maintained by dynamo processes. The activity of partially convective stars is described by the well-studied solar-type $\alpha-\omega$ dynamo \citep{1955ApJ...122..293P,1969AN....291...49S,1982A&A...108..322R,1999ApJ...524..295S,charbonneau2010dynamo}. 

Partially convective stars are expected to have magnetic activity similar to that observed in the Sun. \citet{1972ApJ...171..565S} showed that rotation plays an important role in the generation of stellar activity. They found that the level of magnetic activity increases with increasing rotation (shorter period) until a plateau is reached. This so-called `saturation' of the magnetic activity is known since the early 80ies \citep[e.g.][]{1983ASSL..102..475V,1984A&A...133..117V} and is observed for late type stars in many activity tracers e.g. X-ray luminosity \citep[e.g.][]{2019A&A...628A..41P}, chromospheric emission in the Ca II H\&K lines \citep[e.g][]{1984ApJ...279..763N}, H$\alpha$ emission \citep[e.g.][]{2017ApJ...834...85N}, UV emission \citep[e.g.][]{1987ApJ...316..434S} or photometric variability due to star spots or stellar flares \citep[e.g.][]{2016MNRAS.463.1844S}. Slowly rotating late type stars are characterized by activity cycles similar to the $\sim11$\,yr solar cycle. For fast rotators in the saturated regime, however, activity cycles might be suppressed by the large coverage of the stellar surface with active structures \citep[as shown e.g. in X-rays by ][]{2022AN....34310066C}.

From helioseismic observations it is known that the solar rotation is non-uniform and that the angular velocity declines from equator to pole \citep{2005PhyU...48..449K}. Such differential rotation is observed in many stars and plays a key role in stellar activity \citep{1979cmft.book.....P}; activity cycles, flares, and star spots are related to differential rotation. The amplitude of differential rotation gives information on the type of dynamo operating inside the star. In particular, differential rotation is responsible for generating strong toroidal magnetic fields from weak poloidal fields, which characterizes the $\alpha-\omega$ dynamo \citep{1961ApJ...133..572B}. Hence, measurements of differential rotation are essential to investigate the connections between rotation, convection, and the topology of stellar magnetic fields.

Detailed investigations of magnetic activity, rotation and differential rotation with high precision photometry for a large number of stars started to be in the focus of research 
with the launch of space telescopes like CoRoT \citep[Convection, Rotation, and planetary Transits,][]{2009A&A...506..411A} and \textit{Kepler} \citep{2010Sci...327..977B}. To study activity cycles, however, long-term observations are needed, as the typical time-scales of the cycles range from a few years to decades.

Combining data of the Transiting Exoplanet Survey Satellite \citep[TESS,][]{2014SPIE.9143E..20R} which was launched in 2018 with light curves obtained with \textit{Kepler} and its successor \textit{K2} \citep[Kepler Two-Wheel mission,][]{2014PASP..126..398H} who operated between 2009 and 2017 enables stellar activity studies using high precision photometry on a decade timescale. In particular, combined \textit{Kepler}/\textit{K2} – TESS data provides star spot modulations, rotation rates as well as flare activity levels on a timescale that is comparable to the Sun’s $\sim$11 yr activity cycle. Hence, this long-term photometric monitoring will yield indications of stellar magnetic activity cycles \citep[e.g.][]{2002A&A...393..225M,2019A&A...622A..85N,2019MNRAS.484.3244M}, and help to refine differential rotation \citep{2013A&A...560A...4R,2016A&A...591A..43D}. The unique ability for the combination of \textit{Kepler} and TESS data to detect activity cycles was first demonstrated by \citet{2020AJ....160...36D} who carried out a dedicated study for the very active M4 dwarf GJ\,1243 but could find no sign of solar-like activity cycles over the 10 years baseline. Hence, GJ\,1243 may act as an example of stars in the saturated regime of the rotation-activity-relation that do not show an activity cycle.

Here we present a detailed study of rotation rates, star spot distributions and flare activity for two early M dwarfs, EPIC\,202059229 and EPIC\,245919787, using a combination
of light curves from \textit{K2} and TESS that span a baseline up to $4.5$\,years. Furthermore, our analysis yields limits on activity cycles and differential rotation. We 
motivate the selection of our targets in Sect.~\ref{target_selection} and explain the basic light curve analysis in Sect.~\ref{LC_analysis}. We introduce our methods of the simulation of the spot induced variability amplitudes and the light curve modeling in Sect.~\ref{amplitude_sim} and Sect.~\ref{SOAP2}. The individual results of our analysis for the two selected M dwarfs are summarized in Sect.~\ref{results}. Finally we conclude and provide a brief summary of our results in Sect.~\ref{Conclusions}.


\section{Target selection}
\label{target_selection}

Our study on the activity and rotation of M dwarfs is based on the Superblink proper motion catalog  \citep{2011AJ....142..138L}, an all-sky catalog of $\sim$9000 M dwarf stars with apparent infrared magnitude $J< 10$. In  \citet{2016MNRAS.463.1844S} and \citet{2020AN....341..519R} we presented rotation periods and photometric activity indicators for all 430 M dwarfs from \citet{2011AJ....142..138L} that were observed with the \textit{K2} mission in its 20 campaigns (C0-C19). 

\textit{K2} and  the TESS prime mission followed complementary observing strategies. While the fields observed by \textit{K2} were distributed around the ecliptic plane, TESS performed a near all-sky survey with one of its cameras centered on the ecliptic pole and not covering the ecliptic plane. Hence, the overlap between \textit{K2} and the TESS prime mission is marginal.

From the 430 \citet{2011AJ....142..138L} M dwarfs observed with \textit{K2} only 45 were also covered in the TESS prime mission. A list of these targets with the corresponding TIC IDs is given in Table~\ref{K2_TESS_targets}. The rotation period and flare frequency measured in the \textit{K2} light curves are also listed. These values were extracted from \citet{2020AN....341..519R}.

As shown in \citet{2022A&A...665A..30S} the standard deviation of the light curve after subtraction of rotational modulation and flares (cleaned and flattened light curve), S$_{\mathrm{flat}}$, of the TESS light curves are higher compared to \textit{K2} which is a consequence of the smaller telescopes used by the TESS mission. As a result, low variability amplitudes measured with \textit{K2} will be hidden in the noise of TESS. For our combined \textit{K2} and TESS analysis it is essential that a clear variability is detected in both light curves. Hence, we removed stars from the list that have a lower variability amplitude measured with \textit{K2} than the spread of the TESS cleaned and flattened light curve (minimum flux subtracted from maximum flux). This leaves 19 out of 45 stars that have a full \textit{K2} variability amplitude which can significantly be detected with TESS. 

Not all of these 19 stars show a clear variability in the TESS light curves. Two main reasons are responsible for the lower variability in TESS. Either the full variability amplitude is not covered by the duration of a TESS light curve due to a long rotation period or the level of stellar activity has changed over time. To select targets that show a significant brightness modulation we calculated the reduced $\chi^{2}$ of the TESS light curves. Only stars with a $\chi^{2}_{\mathrm{red}}>1.5$ were considered as variable. Of the resulting 10 stars, one star, EPIC\,246732310, has a rotation period measured with \textit{K2} that exceeds half the observational baseline of TESS. Hence, with TESS the determination of a reliable period is not possible. Further two stars (EPIC\,202059199 and EPIC\,211498244) do not show a periodic signal in their TESS light curves. The high $\chi^{2}_{\mathrm{red}}$ value of these two stars is a result of a large flare and an eclipse for EPIC\,202059199 and EPIC\,211498244, respectively. 

Finally, we excluded all multiple systems from our analysis. Three stars (EPIC\,201909533, EPIC\,202571062, EPIC\,201497396) are listed in the Washington Double Star Catalog \citep{2023yCat....102026M} and two more, EPIC\,201501470 and EPIC\,246625561, were found to be spectroscopic binaries by \citet{2010ApJ...716.1522S} and \citet{2022ApJ...929..169R}, respectively.
As result of the downselection only two stars, EPIC\,202059229 and EPIC\,245919787, are suited for an in-depth analysis and comparison of the \textit{K2} and TESS light curves. A summary of the stellar parameters obtained from the TESS Input Catalog \citep[TIC v8.1,][]{2018AJ....156..102S} for our two targets are given in Table~\ref{stellar_param}.

\begin{table*}
\centering
\caption{Summary of the rotation and flare analysis of the $45$ M dwarfs from \citet{2011AJ....142..138L} that were also observed in the TESS prime mission. The results from the \textit{K2} mission are adopted from \citet{2020AN....341..519R}.} 
\label{K2_TESS_targets}
\begin{tabular}{llr@{\,$\pm$\,}llll}
\hline \hline
EPIC ID	&	CAMPAIGN	&	\multicolumn{2}{c}{$P_{\mathrm{rot,K2}}$ [d]}	&			Nflare/day$_{\mathrm{K2}}$	&	TIC ID	&	Sector	\\ \hline
202059198	&	C0	&	27.307	&	0.465	&	0.030	&	385812015	&	6	\\
202059199	&	C0	&	\multicolumn{2}{c}{---}			&	0.121	&	385811983	&	6	\\
202059222	&	C0	&	72.160*	&		&	---	&	438489075	&	6	\\
202059223	&	C0	&	\multicolumn{2}{c}{---}			&	---	&	438496720	&	6	\\
202059229	&	C0	&	5.006	&	0.025	&	0.121	&	372611670	&	6	\\
201909533	&	C1	&	2.294	&	0.003	&	0.200	&	291074569	&	22	\\
202571062	&	C2	&	3.988	&	0.017	&	0.077	&	392323519	&	12	\\
202748218	&	C2	&	120.610*	&		&	---	&	68559790	&	12	\\
206007536	&	C3	&	16.095	&	0.503	&	0.030	&	188578639	&	2	\\
206019392	&	C3	&	75.440*	&		&	0.225	&	188580272	&	2	\\
210317378	&	C4	&	26.909	&	0.202	&	0.058	&	345454031	&	5	\\
210340480	&	C4	&	\multicolumn{2}{c}{---}			&	0.015	&	311128611	&	5	\\
210360545	&	C4	&	\multicolumn{2}{c}{---}			&	---	&	242985446	&	5	\\
210393283	&	C4	&	16.626	&	0.093	&	---	&	434135656	&	5	\\
210408563	&	C4	&	17.080	&	0.648	&	0.073	&	435881860	&	5	\\
210434433	&	C4	&	23.278	&	0.638	&	0.044	&	348663808	&	5	\\
210434769	&	C4	&	\multicolumn{2}{c}{---}			&	---	&	348663813	&	5	\\
212323908	&	C6	&	\multicolumn{2}{c}{---}			&	---	&	335010451	&	10	\\
213817346	&	C7	&	0.812	&	0.004	&	0.184	&	169148174	&	13	\\
220215093	&	C8	&	57.550	&	0.254	&	0.104	&	399702696	&	3	\\
228711280	&	C10	&	60.700	&	6.070	&	---	&	110996526	&	10	\\
229149393	&	C10	&	8.389	&	0.083	&	---	&	291109924	&	23	\\
242219524	&	C11	&	23.946	&	0.221	&	0.014	&	105559958	&	12	\\
245919787	&	C12	&	5.649	&	0.016	&	0.076	&	434101713	&	2	\\
245983914	&	C12	&	10.792	&	0.366	&	0.076	&	9051707	&	2	\\
246043232	&	C12	&	90.667*	&		&	0.038	&	9633428	&	2	\\
246048874	&	C12	&	19.272	&	0.029	&	0.038	&	9062134	&	2	\\
210491860	&	C13	&	15.031	&	0.158	&	---	&	456892725	&	5	\\
246625561	&	C13	&	2.215	&	0.041	&	0.063	&	436584697	&	5	\\
246732310	&	C13	&	12.933	&	0.047	&	0.352	&	436655312	&	5	\\
246807434	&	C13	&	107.122*	&		&	---	&	436678563	&	5	\\
246862801	&	C13	&	15.992	&	0.080	&	0.038	&	436635454	&	5	\\
201497396	&	C14	&	13.590	&	0.173	&	0.180	&	281746741	&	9	\\
201501470	&	C14	&	1.304	&	0.045	&	0.090	&	281731214	&	9	\\
248425357	&	C14	&	108.485*	&		&	0.026	&	277811911	&	8	\\
248453031	&	C14	&	34.232	&	0.895	&	0.077	&	374312723	&	9	\\
249101965	&	C15	&	66.301	&	1.081	&	0.012	&	185856339	&	11	\\
249186244	&	C15	&	\multicolumn{2}{c}{---}			&	---	&	186072161	&	11	\\
249338840	&	C15	&	22.354	&	0.081	&	0.069	&	70111746	&	11	\\
251550724	&	C17	&	46.965	&	0.134	&	0.107	&	115070878	&	23	\\
251567386	&	C17	&	\multicolumn{2}{c}{---}			&	---	&	130718008	&	23	\\
251583820	&	C17	&	52.926	&	0.256	&	---	&	66765158	&	23	\\
251584738	&	C17	&	41.648	&	1.080	&	0.015	&	66766055	&	23	\\
211385897	&	C5,C18	&	11.508	&	0.298	&	0.211	&	366499976	&	7	\\
211498244	&	C5,C18	&	\multicolumn{2}{c}{---}			&	0.008	&	27654301	&	7	\\
\hline
\end{tabular}
\\
*the period exceeds the duration of the \textit{K2} observation and is, hence, not reliable. No uncertainties are given

\end{table*}

\begin{table}
\centering
\caption{Stellar parameters from the TIC \citep[v8.1,][]{2018AJ....156..102S} for our two selected targets that were observed by \textit{K2} and TESS. The limb darkening (LD) coefficients, $u$ and $v$, for the quadratic limb darkening law were inferred from the tables by \citet{2010A&A...510A..21S} and \citet{2017A&A...600A..30C} (see Sect.~\ref{SOAP2}).}
\label{stellar_param}
\begin{tabular}{lr@{\,$\pm$\,}lr@{\,$\pm$\,}l}
\hline \hline
& \multicolumn{2}{c}{EPIC\,202059229}  & \multicolumn{2}{c}{EPIC\,245919787} \\ 
& \multicolumn{2}{c}{TIC\,372611670} & \multicolumn{2}{c}{TIC\,434101713}\\ \hline
Parameter & \multicolumn{4}{c}{Value}\\ \hline
K2 campaign &  \multicolumn{2}{c}{C00}  & \multicolumn{2}{c}{C12} \\
$K_{\rm p}$ [mag]$^{a}$ &  \multicolumn{2}{c}{10.500} & \multicolumn{2}{c}{10.329}\\
TESS sector & \multicolumn{2}{c}{6} & \multicolumn{2}{c}{2}\\
$T$ [mag] & 9.2238 & 0.0075 & 8.8813 & 0.0069 \\
$R_{\mathrm{\ast}}$ [R$_{\odot}$] & 0.5498 & 0.0163 & 0.8043 & 0.0937 \\
$M_{\mathrm{\ast}}$ [M$_{\odot}$] & 0.545 & 0.020& 0.590 & 0.083 \\
$T_{\mathrm{eff}}$ [K] & 3720 & 157 & 3821 & 123\\
log\,$g$  & 4.6942 & 0.0096 & 4.3981 & 0.1388 \\
Distance [pc] & 22.3819 & 0.0263 & 27.3518 & 0.04365 \\
$u_{\mathrm{K2}}$ & \multicolumn{2}{c}{0.3677} & \multicolumn{2}{c}{0.4515} \\
$v_{\mathrm{K2}}$  & \multicolumn{2}{c}{0.3552} & \multicolumn{2}{c}{0.3000} \\
$u_{\mathrm{TESS}}$   & \multicolumn{2}{c}{0.2763} & \multicolumn{2}{c}{0.3667} \\
$v_{\mathrm{TESS}}$  & \multicolumn{2}{c}{0.3605} & \multicolumn{2}{c}{0.2858} \\
Spectral type$^{b}$ & \multicolumn{2}{c}{M1.0} & \multicolumn{2}{c}{M1.0} \\
\hline
\end{tabular}
\\
$^{a}$from the K2 Ecliptic Plane Input Catalog \citep[EPIC,][]{2017yCat.4034....0H}\\
$^{b}$from \citet{2020AN....341..519R}
\end{table}

\section{Basic light curve analysis}
\label{LC_analysis}

For our two selected targets we analyzed in total four light curves, one \textit{K2} and one TESS light curve for each star. The \textit{K2} and TESS light curves were downloaded from the  MAST (Barbara A. Mikulski Archive for Space Telescopes) Portal.

Both targets were observed with \textit{K2} in long ($\sim$30\,min data-point cadence) and short ($\sim$1\,min data-point cadence) cadence mode. For the long cadence data we made use of the light curve produced by the EPIC Variability Extraction and Removal for Exoplanet Science Targets pipeline \citep[EVEREST,][]{2016AJ....152..100L}. We complemented the obtained results with our previous analysis from \citet{2020A&A...637A..22R} for the short cadence light curves.
 
For the TESS data we inspected the Simple Aperture Photometry (SAP) light curve as well as the Pre-search Data Conditioning Simple Aperture Photometry (PDCSAP) light curve. If the TESS detrending pipeline was able to remove all obvious systematic trends we directly used the PDCSAP light curves. Otherwise we had to apply an additional  light curve treatment (see Sect.~\ref{EPIC202059229_rotation}). In the TESS light curves we removed all data points with an assigned quality flag different from 0 except of `Impulsive outlier' (which could be real stellar flares) and `Cosmic ray in collateral data' \citep[see more detailed information in ][]{2022A&A...661A..29M}. 

The TESS and the \textit{K2} light curves were then normalized by dividing all data points by the median flux. In those final light curves we have determined the rotation period, variability amplitudes and flare frequency for all light curves using the same methods. These methods are summarized in the following. 

\subsection{Rotation period and variability amplitude}
\label{period}

To measure the rotation period, we initially applied three different standard time series analysis techniques, a generalized Lomb-Scargle periodogram \citep[\begin{small}GLS\end{small};][]{2009A&A...496..577Z}, the autocorrelation function (ACF), and fitting the light curves with a sine function. The period search method is described in detail in our previous works, e.g. \citet{2020A&A...637A..22R}. 

To provide an even more precise rotation period measurement we applied a fourth method. We optimized the period through phase-folding of the light curve with 10000 test periods changed in steps of 0.00001\,d around an initial period which was chosen as a less precise value of the initial period estimate (one digit after the comma).  A smoothed light curve was then subtracted from each phase-folded light curve. The period that resulted in the lowest standard deviation of the residuals was finally chosen as the final adopted rotation period. 

For the determination of the uncertainties we calculated the standard deviation of the period values from the four different methods. In addition, we used the formulas given in \citet{1985PASP...97..285G} to calculate an error for the rotation period. For each star, we adopted either the standard deviation or the calculated error as the final uncertainty, whichever was the higher value.

Visual inspection of the phase-folded light curves indicate a change of the variability amplitude of the spot induced brightness modulation, $A_{\mathrm{var}}$, for our targets. We define $A_{\mathrm{var}}$ as the peak-to-peak difference of the relative flux from the maximum to the primary minimum. The latter is the deepest minimum in the double-humped light curve in a single rotation. This variation is caused by star spots that are moving and evolving on time scales shorter than the length of the individual observation. To monitor the changes of $A_{\mathrm{var}}$ over time we determined it individually for each rotation. We obtained the time and the flux values of each minimum and maximum in the light curve using the IDL routine \begin{small}MPFITPEAK\end{small}. $A_{\mathrm{var}}$ was then calculated as the flux difference between neighboring extrema. In order to associate one amplitude value per light curve we calculate the mean $A_{\mathrm{var}}$ by taking the average, $\bar{A}_{\mathrm{var}}$. The uncertainty of $\bar{A}_{\mathrm{var}}$ is defined as the standard deviation of the individual amplitudes. The final adopted values for the variability amplitudes for our two targets are given in Table~\ref{amplitudes}.

\subsection{Flares}
\label{flares}

The algorithm for the detection of stellar flares that is based on the routine developed by \citet{2016MNRAS.463.1844S} was explained in detail by \citet{2020A&A...637A..22R} and \citet[][application to TESS data]{2022A&A...665A..30S}. In short, we create a smoothed light curve with an iterative process of boxcar smoothing which is then subtracted from the original light curve to remove the rotational signal, thus generating a flat light curve. All groups of at least five (for \textit{K2} short cadence) or three (for TESS) consecutive data points of the original light curve that lie at least 3$\sigma$ above the final flattened light curve are flagged as potential flares. Then, we applied five criteria in order to validate these potential flares. These criteria are: (1) The flare event must not occur right before or after a gap in the light curve; (2) the flux ratio between the flare peak and the last flare point must be $\geq$2; (3) the flare peak can not be the last point of the potential flare; (4) the decay phase must be longer than the rise phase; (5) a fit conducted using the flare template defined by \citet{2014ApJ...797..122D} must fit the flare better than a linear fit through the same flare points.

The flare energy in the instrument's waveband (\textit{Kepler} or TESS, respectively) is computed following the approach of \citet{2016ApJ...829...23D}. The equivalent duration{, defined as the amount of time that it would take a star, in its quiescent state, to release the same amount of energy released during a flare \citep{1972Ap&SS..19...75G, 2012PASP..124..545H}, which is computed as the integral under the flare, is multiplied by the quiescent stellar luminosity which we obtained by converting the $K_{\rm p}$ and $T$ magnitudes to fluxes using the zero-points and effective wavelengths provided at the filter profile service of the Spanish Virtual Observatory \citep[SVO,][]{2012ivoa.rept.1015R} and applying the {\it Gaia}-DR2 distances given in Table~\ref{stellar_param}. Finally, in order to compare flare rates between \textit{K2} and TESS we compiled the cumulative flare frequency distributions \citep[FFD,][]{1976ApJS...30...85L,2013ApJS..209....5S} and we computed a specific flare rate $\nu$, a frequency for flares of a given energy or larger as defined by \citet{2019ApJ...871..241D}, for our targets.

\section{Simulation of the amplitudes of spot induced variability}
\label{amplitude_sim}

If the stellar surface is covered by spots the observed spectrum is a combination of the photospheric spectrum and the spectrum of the spots. In general, spots can be cooler or hotter than the photosphere.

The variability amplitude, $A_{\mathrm{var}}$, is wavelength dependent and is a function of the difference between effective temperature and spot temperature, $T_{\mathrm{eff}}-T_{\mathrm{spot}}$ and the spot size, $R_{\mathrm{spot}}(R_{\mathrm{*}})=\sqrt{2f}$, where $f$ is the spot filling factor which is defined as $f=Area_{\mathrm{spot}}/Area_{\mathrm{*}}$ where $Area_{\mathrm{*}}$ is the area of the visible hemisphere of the star. Hence, measuring the variability amplitude in different filters yields constraints on the size and the temperature of star spots. 

We applied the method described by \citet{2005A&A...438..675S} to model the amplitudes of photometric light curves. In particular, we calculated a grid of amplitudes for a range of values for the parameters $f$ and $T_{\mathrm{spot}}$. 

For the calculation of the amplitudes we used the NextGen model atmospheres \citep{1997ARA&A..35..137A,1997A&A...327.1054B,1998A&A...337..403B,1999ApJ...512..377H} which we downloaded from the Theoretical spectra web server of the Spanish Virtual Observatory \citep[SVO,][]{2006ASPC..351...19G}. 
Analogous to \citet{2005A&A...438..675S} we call the spectrum of the spotless photosphere $S_{0}$ and the spectrum of the spots $S_{\mathrm{spot}}$. For the simulation of the amplitudes we assume that there is one cool spot with a certain temperature and size on the surface of the star. The stellar spectrum representing photosphere and spot at the light curve minimum when the spot is on the visible hemisphere of the star, $S_{\mathrm{min}}$, is then defined as
\begin{equation}
   S_{\mathrm{min}}=(1-f) \times S_{0}+f \times S_{\mathrm{spot}}(T_{\mathrm{spot}}).
\end{equation}
The spectra $S_{0}$ and  $S_{\mathrm{min}}$ were convolved with the transmission curves of \textit{Kepler} and TESS which we downloaded from the filter profile service of the SVO \citep{2012ivoa.rept.1015R} to compute the fluxes in the two bands, $F_{\mathrm{0,K2}},F_{\mathrm{min,K2}}$ and $F_{\mathrm{0,TESS}},F_{\mathrm{min,TESS}}$.
Finally, we calculated the amplitudes as
\begin{equation}
\label{amplitude_formular}
    A_{\mathrm{X}}=-2.5\,\mathrm{log}\frac{F_{\mathrm{min,X}}}{F_{\mathrm{0,X}}},
\end{equation}
where X is the \textit{Kepler} or TESS filter and $F_{\mathrm{0,X}}$ the flux of the spotless photosphere in the respective wavelength band. An example for the method is given in Fig.~\ref{example_spectra} for a fictitious star with $T_{\mathrm{eff}}=3000$\,K, log\,$g=5.0$, and $R_{\mathrm{\ast}}=0.20$\,R$_{\odot}$ at a distance of 14\,pc. In that example, as spot temperature and size we used $T_{\mathrm{spot}}=2500$\,K and $f=0.1$. As a result we obtained amplitudes of 5.96\% and 6.73\% for \textit{Kepler} and TESS, respectively. The computed amplitudes for the same assumed spot with given size and contrast is only slightly different in the two filters. Hence, we can compare variability amplitudes measured in \textit{Kepler} and TESS light curves,  and any detected changes can be assumed to be a sign of a changing level of stellar activity.

\citet{2005A&A...438..675S} used such simulations to determine spot temperatures and spot sizes for data taken quasi-simultaneously in different bandpasses. The observations of our targets by \textit{Kepler} and TESS are, however, separated by several years and, hence, spots may have evolved, disappeared and re-appeared. Therefore, it is not possible to find only one solution with a unique spot temperature and size. The aim of this calculation is to determine the input parameters for the light curve modeling (see Sect.~\ref{SOAP2}). Therefore, we treated the \textit{K2} and TESS light curves individually. As a consequence of the non-simultaneous observations, the amplitude calculation turned into a degenerate problem, i.e. a lower $T_{\mathrm{spot}}$ with lower $f$ results in the same amplitude  as a higher $T_{\mathrm{spot}}$ with a higher $f$. \citet{2005LRSP....2....8B} showed that the spot temperature contrast ($T_{\mathrm{eff}}-T_{\mathrm{spot}}$) with respect to $T_{\mathrm{eff}}$ can be represented by a second order polynomial. From this result \citet{2019ApJ...876...58N} deduced a relation for the temperature difference between photosphere and spots (their Eq. 4).
We used that relation to constrain $T_{\mathrm{spot}}$ and to resolve the degeneracy. We then calculated the spot induced variability amplitudes using Eq.~\ref{amplitude_formular} for the constrained $T_{\mathrm{spot}}$ and a grid of $f$. In the initial run a $f$ between 0.01 and 0.5 in steps of 0.01 was used. In the final run we refined our grid by using a step size of 0.001 around the best-fitting model obtained in the initial run.

\begin{figure}
  \centering
  \includegraphics[width=0.3\textwidth,angle=270]{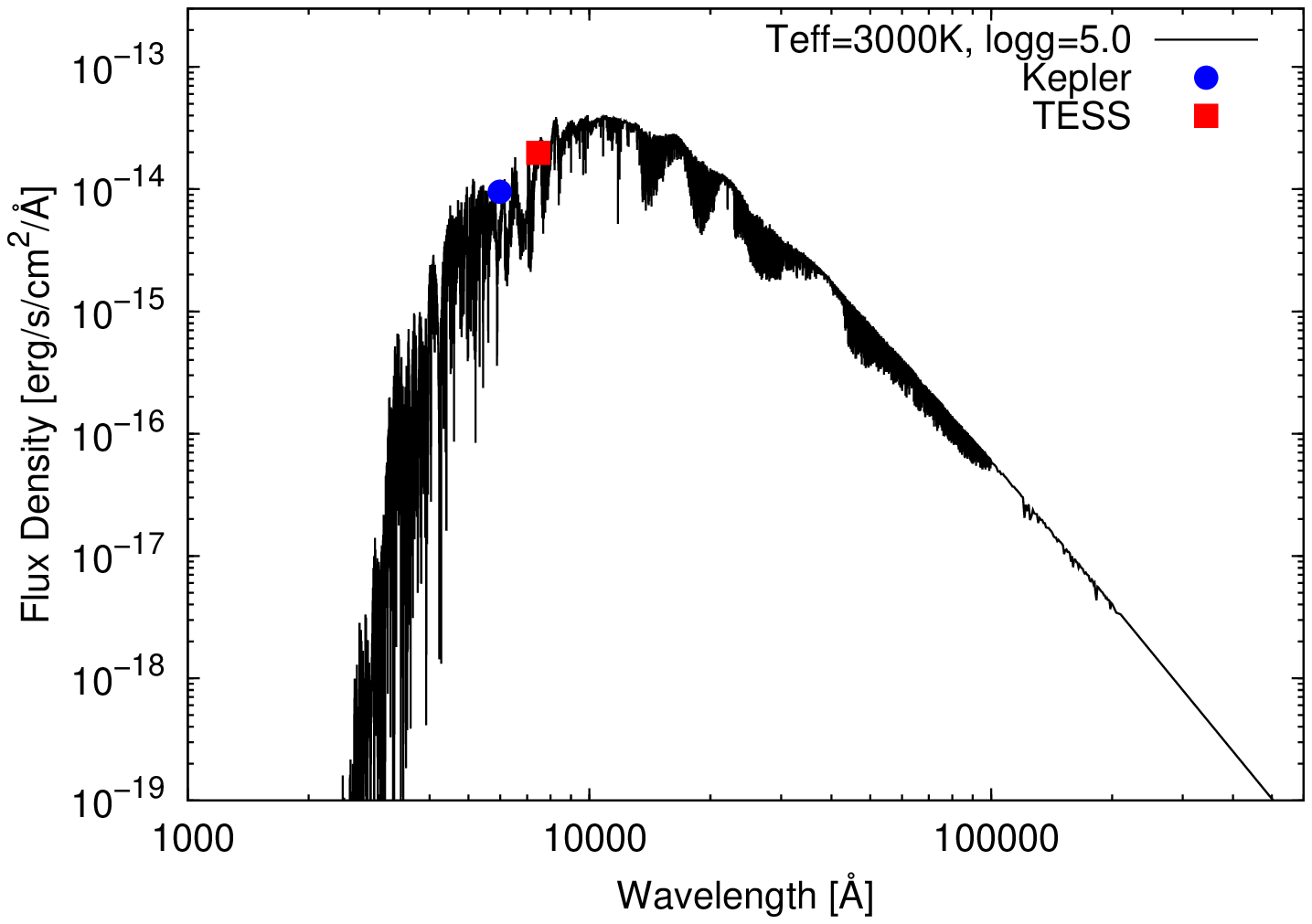}
  \includegraphics[width=0.3\textwidth,angle=270]{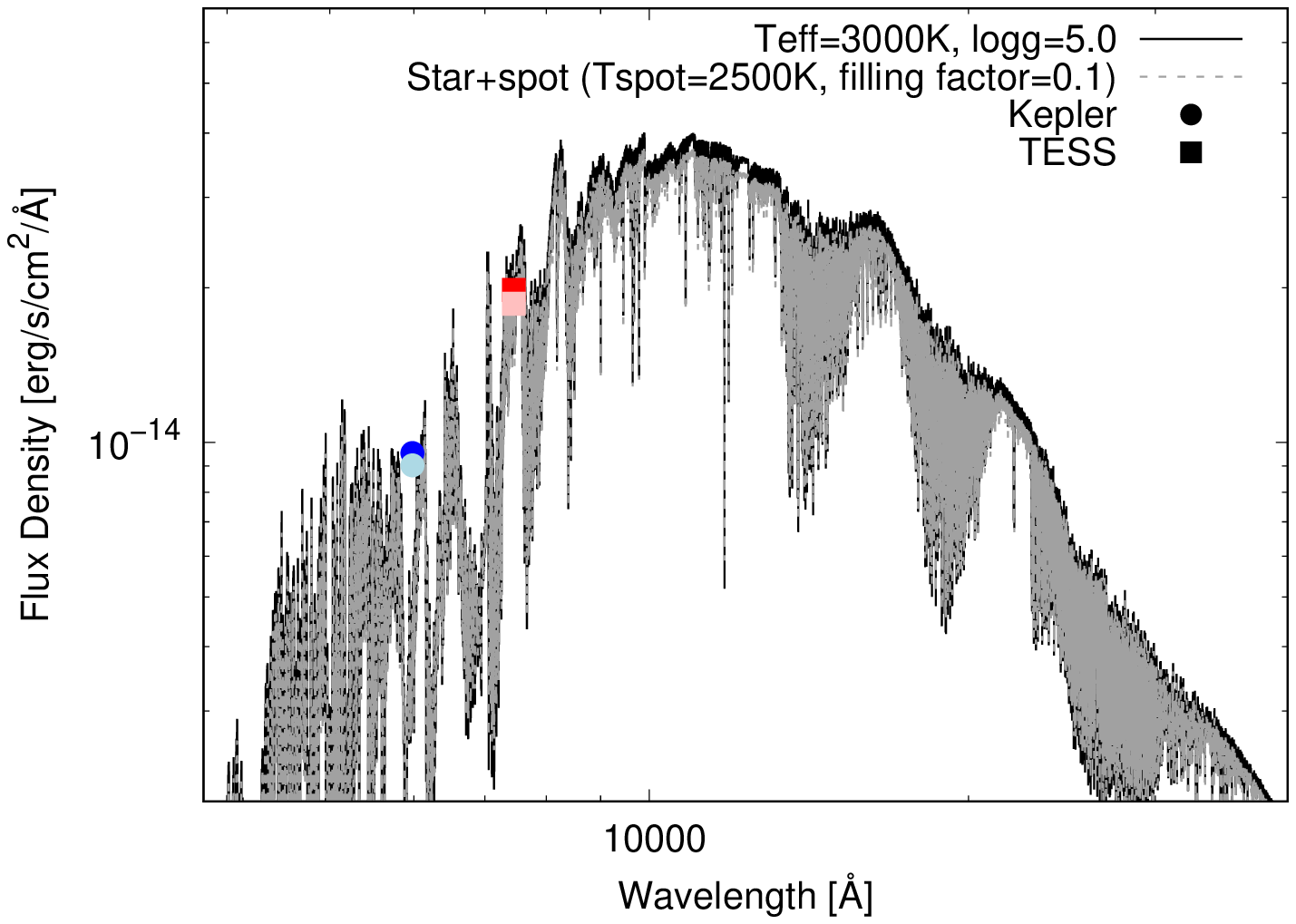}
  \caption{Example spectra for a fictitious star ($T_{\mathrm{eff}}=3000$\,K, log\,$g=5.0$, $R_{\mathrm{\ast}}=0.20$\,R$_{\odot}$, distance\,=\,14\,pc). Top: Spectrum of the spotless photosphere. The blue circle and the red square mark the fluxes in the \textit{K2} and TESS wavelength band, respectively. Bottom: Zoom into the spectrum. The spectrum of the spotted photosphere is overplotted as dashed gray line. The parameters of the spot are $T_{\mathrm{spot}}=2500$\,K and $f=0.1$. he fluxes in the \textit{K2} and TESS wavelength band that include the star spot are shown as light blue square and pink circle, respectively. The resulting, spot modulation amplitudes in the light curve are 5.96\% for \textit{K2} and 6.73\% for TESS.}
  \label{example_spectra}
\end{figure}

\section{Light curve modeling}
\label{SOAP2}

For the modeling of the spots in the \textit{K2} and TESS light curves we used \begin{small}SOAP 2.0\end{small} \citep{2014ApJ...796..132D}, an improved version of the Spot Oscillation And Planet \citep[\begin{small}SOAP\end{small},][]{2012A&A...545A.109B} code. \begin{small}SOAP 2.0\end{small} includes (besides the flux effects of active regions to estimate the activity-induced photometric variation) also additional stellar physics, namely the convective blueshift inside active regions, the limb brightening effect of plages, a two parametric quadratic limb darkening law, and a realistic spot and plage contrast ratio \citep{2014ApJ...796..132D}. It is, hence, more realistic than other tools that only consider the stellar flux.

\begin{small}SOAP 2.0\end{small} models the surface of a rotating star with up to four different active regions and returns the resulting light curve. The necessary input stellar parameters are the radius, the rotation period, the stellar inclination angle, the effective temperature of the star, the difference in temperature between the star effective temperature and the spot temperature, and the two limb darkening coefficients. Furthermore, the longitude, the latitude and the size of each active region are required. Since \begin{small}SOAP 2.0\end{small} does not fit the light curve but produces a model for given parameters we created a grid of models and compared them to the observations. However, having at least ten different input parameters (seven stellar parameters and three parameters per active region) increased the computing time immensely. Moreover, spot modeling is a highly degenerate problem, meaning that changing different parameters, e.g. the stellar inclination angle and the spot latitude, can result in identical light curve models. Hence, a physical spot model can only be obtained with additional information from other types of observations, e.g. spectral analysis. More information on the degeneracy of the models is given in Appendix~\ref{degeneracy}. 

Since we do not have access to additional information other than the photometric data, we had to make some assumptions for feasible modeling. In all our models we fixed the stellar inclination angle (to $90$$^{\circ}$; equator on), the radius, the effective temperature, the beforehand determined rotation period and the limb darkening coefficients. The theoretical limb darkening coefficients for the quadratic law (given in Table~\ref{stellar_param}) were bilinearly interpolated (in effective temperature and surface gravity) from the tables by \citet{2010A&A...510A..21S} and  \citet{2017A&A...600A..30C} for the \textit{K2} and the TESS observations, respectively, using the stellar parameters given in the TIC (see Table~\ref{stellar_param}). Furthermore, we assumed two dark spots (justified by the double-humped light curve shape) of the same size and that the difference in the amplitudes of the primary and the secondary minimum is purely caused by the different location on the stellar surface. The life-time of the spots is assumed to be longer than the duration of the individual \textit{K2} and TESS observation. The spot contrast and size, that we assume to be the same for both spots, were taken from the amplitude simulation as explained in Sect.~\ref{amplitude_sim}. Finally, the only free parameters in the model were the longitude and the latitude of the stellar spots.

We searched for spots on the full star, i.e. the parameter space for the spot position was defined to be longitude between 0$^{\circ}$ and 360$^{\circ}$ and latitude from -90$^{\circ}$ to 90$^{\circ}$. We started with a coarse grid to narrow down the approximate position and then refined the grid to a step size of 0.5$^{\circ}$ in longitude and latitude in the final simulation. 

\begin{small}SOAP 2.0\end{small} is not able to account for changing spots on a stellar surface. However, as mentioned above, the light curves show significant changes of the variability amplitude across the observations. Therefore, we splitted the light curves into several parts to create separate models for each rotation. These light curves parts were chosen to be a little bit longer than one single rotation to allow for an overlap. The individual parts of the model light curve were then combined to retrieve a full model for the \textit{K2} and TESS light curves. The overlap of the individual models allowed us to connect them at the points where they intersect. With this approach we can infer how the spots are moving across the stellar surface during the observations.

\section{Results for individual stars}
\label{results}

In this section we describe the results of the analysis for our two targets EPIC\,202059229 (Sect.~\ref{EPIC202059229_results}) and EPIC\,245919787 (Sect.~\ref{EPIC245919787_results}). 
Since one star is a field M dwarf and the other one belongs to a young moving group they represent two examples of different characteristics of stellar activity.

\subsection{EPIC\,202059229 / TIC\,372611670}
\label{EPIC202059229_results}

EPIC\,202059229 (TIC\,372611670, PM\,J06467+1557, TYC\,1330-879-1) is a M1.0V dwarf that was observed with \textit{K2} in campaign C0 for $\sim$36\,d. It is known as an X-ray source from \textit{ROSAT} observations \citep[e.g.][]{2003A&A...403..247F}. Its photometric variability was first studied with observations from the All Sky Automated Survey (ASAS) where \citet{2012AcA....62...67K} determined a rotation period of 5.084\,d. Spectroscopic observations with \textit{CARMENES} showed that EPIC\,202059229 has weak $H_{\alpha}$ emission ($H_{\alpha}=-0.61\,\AA{}$). Since it was observed with \textit{K2} in short cadence mode the rotation and photometric activity could be studied in detail by \citet{2015MNRAS.449.3015R} and \citet{2020A&A...637A..22R}. The former authors found that despite the fact that the weak $H_{\alpha}$ emission would have classified it as inactive, the star shows flares roughly once every two days. They obtained a rotation period of 5.04\,d. The power-law slope they found for the FFD ($\beta=-0.69$, read from their Fig.~5) agrees with \citet{2020A&A...637A..22R} within their error bars.

EPIC\,202059229 was observed in the TESS prime mission in sector 6 for $\sim$22\,d with a 2-min cadence. The observations of \textit{K2} and TESS are separated by $\sim$4.5\,yrs and, hence,  the stellar activity level could have undergone changes. The results of the analysis of the \textit{K2} and TESS data with the comparison of the stellar activity are described in the following sections. All stellar parameters of EPIC\,202059229 used in our analysis are summarized in Table~\ref{stellar_param}.

\subsubsection{Rotation period and variability amplitude}
\label{EPIC202059229_rotation}

As described in Sect.~\ref{LC_analysis}, EPIC\,202059229 was observed with \textit{K2} in long and short cadence mode. To save computing time we decided to use only the long cadence data for the period search (and the spot modeling, see Sect.~\ref{EPIC202059229_spots}).

For the TESS data a comparison of the PDCSAP light curve to the SAP light curve revealed that at different times systematic effects were overcompensated by the TESS detrending pipeline and, hence, caused systematic errors (namely flux jumps) that would distort our measurement of the variability amplitude. Therefore, we decided to instead use the SAP data in which we additionally removed a simple second-order polynomial trend after the light curve normalization.

Fig.~\ref{best_Phasfolded_LC_0202059229} shows the phase-folded \textit{K2} and TESS light curves using the refined rotation period as described in Sect.~\ref{period}. Fig.~\ref{Phasfolded_LC_0202059229_K2}, Fig.~\ref{Phasfolded_LC_0202059229_TESS} in Appendix~\ref{period_plots} give the results of the initial period search. The values of the rotation period ($P_{\mathrm{rot,K2}}=5.036\pm0.013$, $P_{\mathrm{rot,TESS}}=4.994\pm0.030$) are consistent within (overlapping) error bars.

\begin{figure}
  \centering
  \includegraphics[width=0.35\textwidth,angle=270]{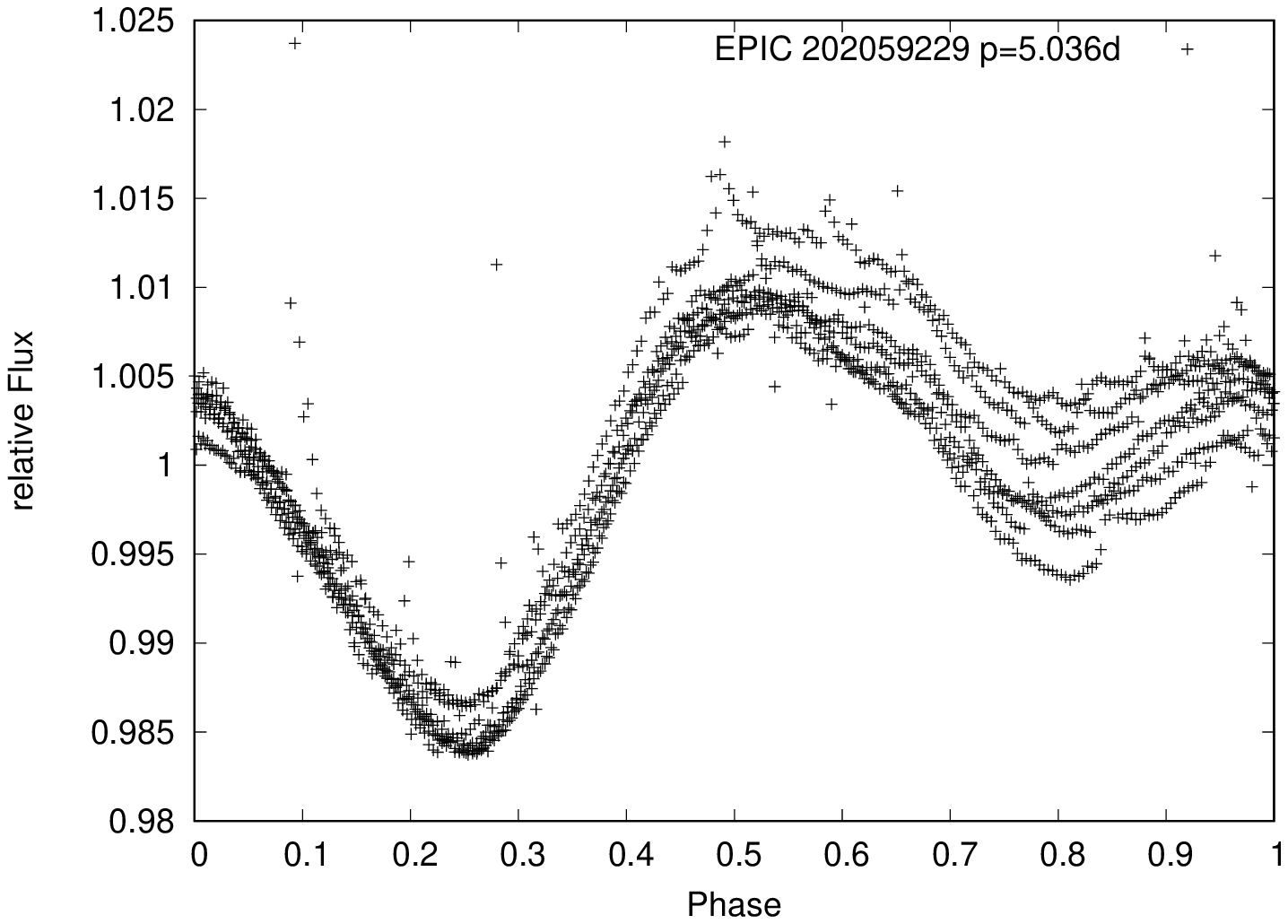}
  \includegraphics[width=0.35\textwidth,angle=270]{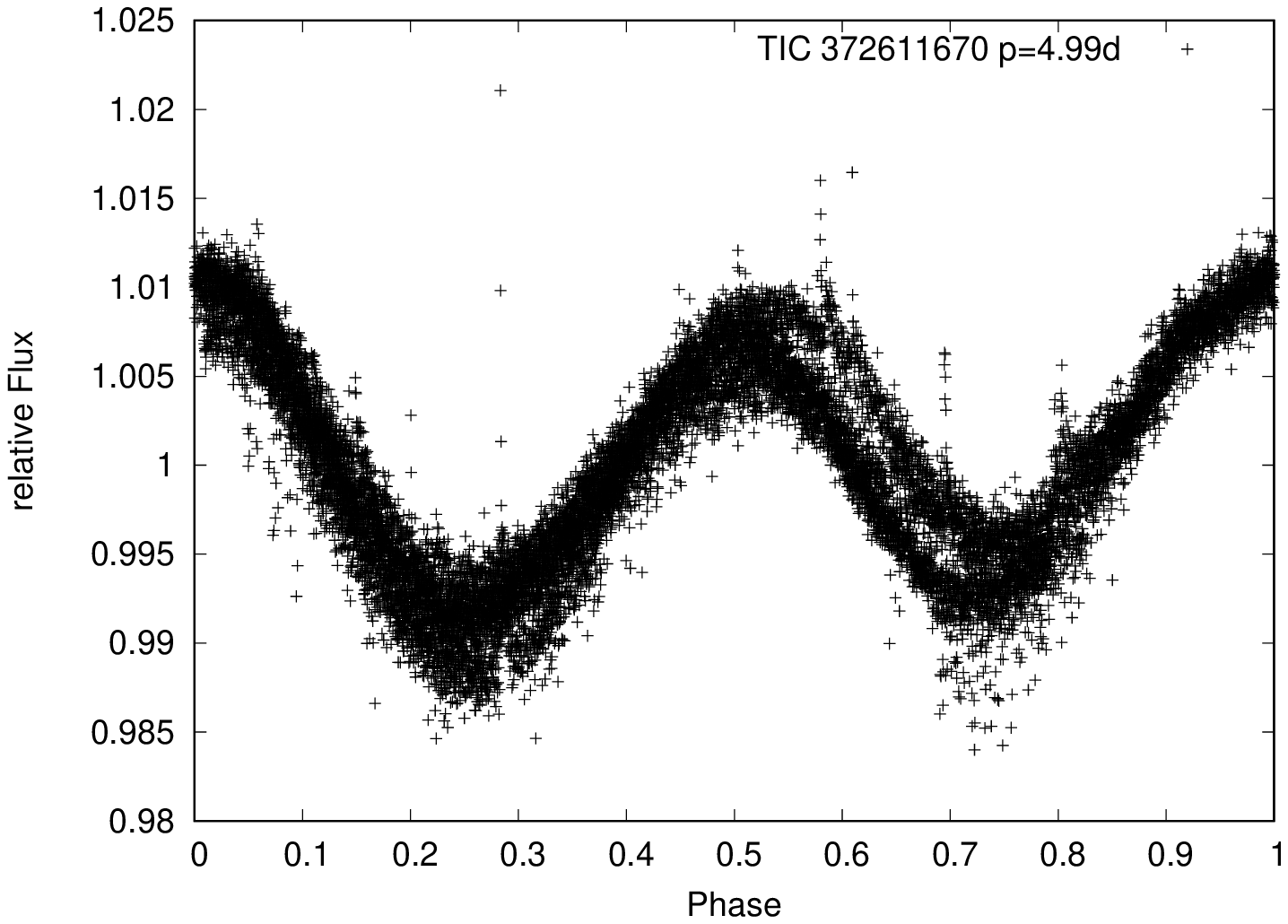}  
  \caption{Light curves of EPIC\,202059229 folded with the refined rotation period (see Sect.~\ref{period}). \textbf{Top:} Light curve observed with \textit{K2} in campaign C0  \textbf{Bottom:} Light curve observed with TESS in sector 6.}
  \label{best_Phasfolded_LC_0202059229}
\end{figure}

As explained in Sect.~\ref{period} we determined $A_{\mathrm{var}}$ individually for each rotation. The data for EPIC\,202059229 yielded seven and four values of $A_{\mathrm{var}}$ for \textit{K2} and TESS, respectively. For one additional cycle in the TESS data we can only give a lower limit, since the maximum is not covered by the data. Fig.~\ref{EPIC202059229_time_ampl} shows the amplitudes over time together with $\bar{A}_{\mathrm{var}}$ that we used as an input for the spot modeling (see Sect.~\ref{EPIC202059229_spots}). The above mentioned lower limit of one of the $A_{\mathrm{var}}$ value entered with a lower weight into the determination of $\bar{A}_{\mathrm{var,TESS}}$. Almost all $A_{\mathrm{var}}$ values for TESS are smaller compared to the values from \textit{K2} except for one which is nearly identical (difference between the two amplitudes is only $3\cdot10^{-5}$). The mean value in the TESS light curve, $\bar{A}_{\mathrm{var,TESS}}=0.0186 \pm 0.0012$, is $\sim 1.3$ times smaller than the \textit{K2} value, $\bar{A}_{\mathrm{var,K2}}=0.0245 \pm 0.0011$. 

In both light curves $A_{\mathrm{var}}$ shows a systematic increase. This behavior could mark an activity cycle. The time between the two almost identical values of $A_{\mathrm{var}}$ is 1717.5\,d which could be a multiple of a putative activity cycle. However, the small number of 12 data points that only cover $\sim$36\,d and $\sim$22\,d for\textit{K2} and TESS, respectively, and a total baseline of $\sim$4.5\,yrs is not sufficient to determine a cycle period $P_{\mathrm{cyc}}$. 

\begin{figure}
  \centering
  \includegraphics[width=0.35\textwidth,angle=270]{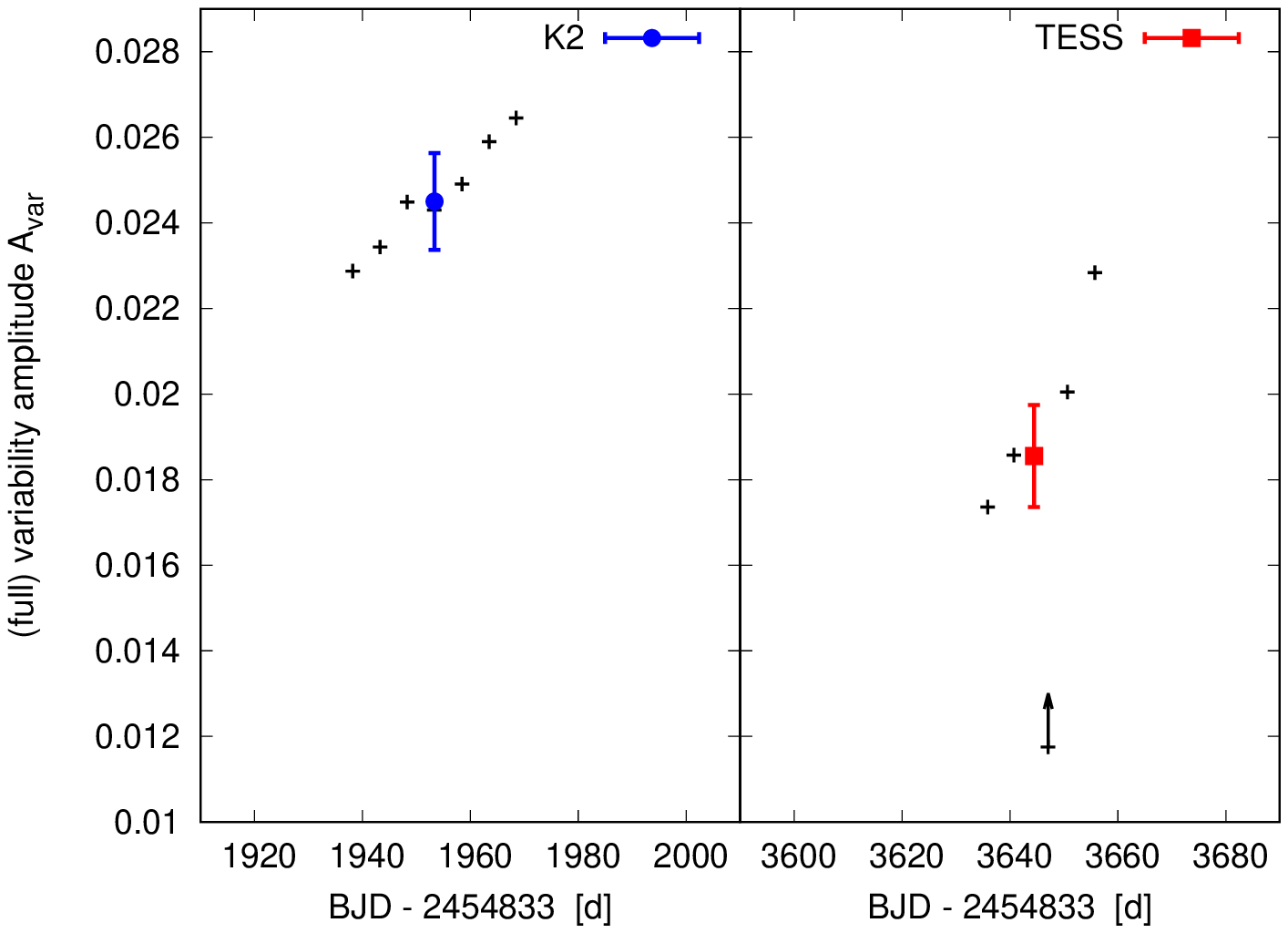}
  \caption{Variability amplitudes measured in the \textit{K2} and TESS light curves of EPIC\,202059229. The black arrow denotes a lower limit for the amplitude. The blue circle and the red square mark mean values $\bar{A}_{\mathrm{var}}$ that were used in the spot modeling.}
  \label{EPIC202059229_time_ampl}
\end{figure}

Fig.~\ref{EPIC202059229_time_phase_flux} shows the light curves of EPIC\,202059229 in terms of phase over time with the relative flux as color code. In this representation of the light curves it can easily be seen that neither the variability amplitude nor the rotation period and, hence the rotation phase has changed strongly. By displaying only very moderate differences in the spot variability pattern of the \textit{K2} and TESS light curves, we can conclude, that the level of activity was stable over the $\sim$4.5\,yrs between the two observations.

\begin{figure*}
  \centering
  \includegraphics[width=0.35\textwidth,angle=270]{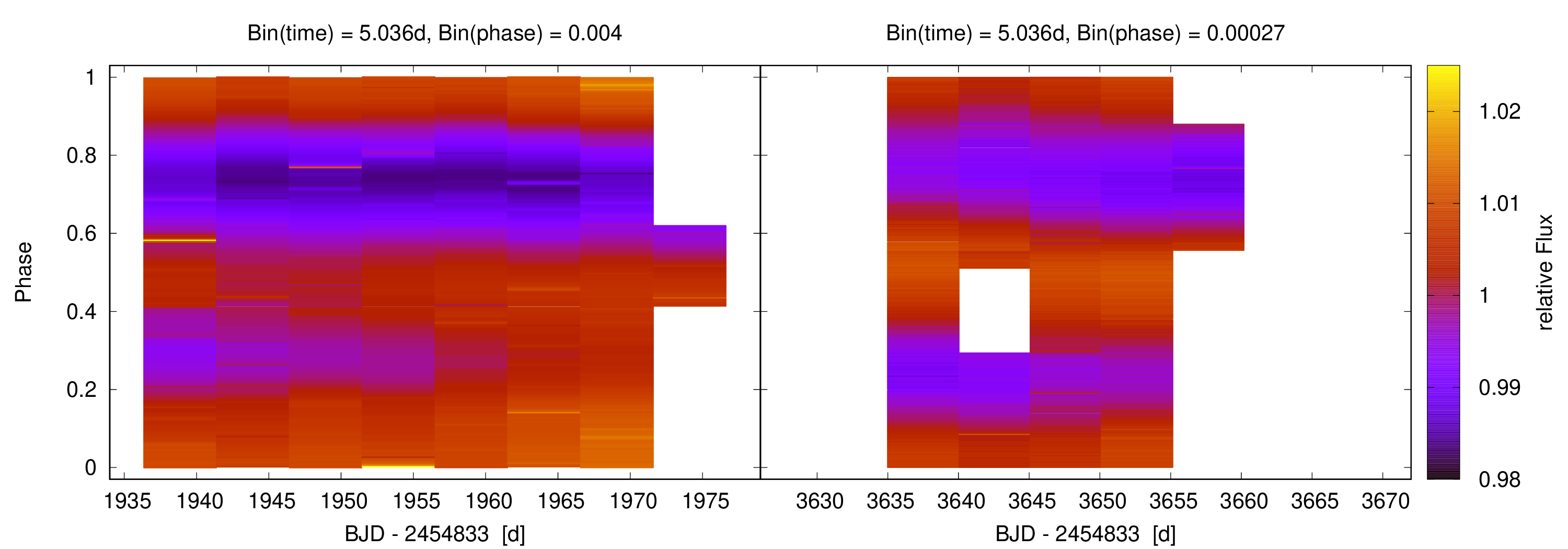}
  \caption{Light curves of EPIC\,202059229 in terms of phase over time with the relative flux as color code. \textbf{Left:} Observations of \textit{K2}. The light curve was splitted into the $\sim$7 rotation cycles present in the total light curve. \textbf{Right:} Observations of TESS. The TESS data only consists of $4.5$ rotation cycles. Both panels were plotted in the same time range and the same color bar to show the light curves on the same scale. The size of the time bins is fixed on the rotation period determined from the \textit{K2} observation ($P_{\rm rot} = 5.036$\,d). The difference in the size of the phase bins is caused by the different observing cadences.}
\label{EPIC202059229_time_phase_flux}
\end{figure*}

\subsubsection{Light curve modeling}
\label{EPIC202059229_spots}

The mean variability amplitudes $\bar{A}_{\mathrm{var,K2}}$ and $\bar{A}_{\mathrm{var,TESS}}$ determined above were used to obtain the spot filling factor ($f$) for the two epochs as explained in Sect.~\ref{amplitude_sim}. The spot temperature of $T_{\mathrm{spot}}=3106$\,K used in the simulation was calculated with Eq.~4 from \citet{2019ApJ...876...58N} inserting $T_{\mathrm{eff}}=3720$\,K (see Table~\ref{stellar_param}). For this value of $T_{\mathrm{spot}}$ we calculated a grid with 50 different filling factors, $f$. 

The initial run of the model revealed that $f$ differs by only $1$\,\% between the \textit{K2} and the TESS observation. Therefore, only a single run of the amplitude modeling with the refined grid was necessary to recover the variability amplitude of both, the \textit{K2} and the TESS observations. Fig.~\ref{EPIC202059229_spot_Grid} shows all models that fit either the \textit{K2} or the TESS observations as the blue and red shaded areas, respectively. For each observation we found seven models from our grid that reproduce the measured value of $\bar{A}_{\mathrm{var}}$ within the error bars. The resulting filling factors $f_{\mathrm{K2}}=0.045 - 0.051$ for \textit{K2} and  $f_{\mathrm{TESS}}=0.034 - 0.040$ for TESS translate into $R_{\mathrm{spot,K2}}= (0.30 - 0.32)\,R_{\mathrm{*}}$ and  $R_{\mathrm{spot,TESS}}= (0.26 - 0.28)\,R_{\mathrm{*}}$. A summary of the rotation periods, amplitudes and the derived spot parameters for both light curves is given in Table~\ref{amplitudes}.

\begin{figure}
  \centering
  \includegraphics[width=0.35\textwidth,angle=270]{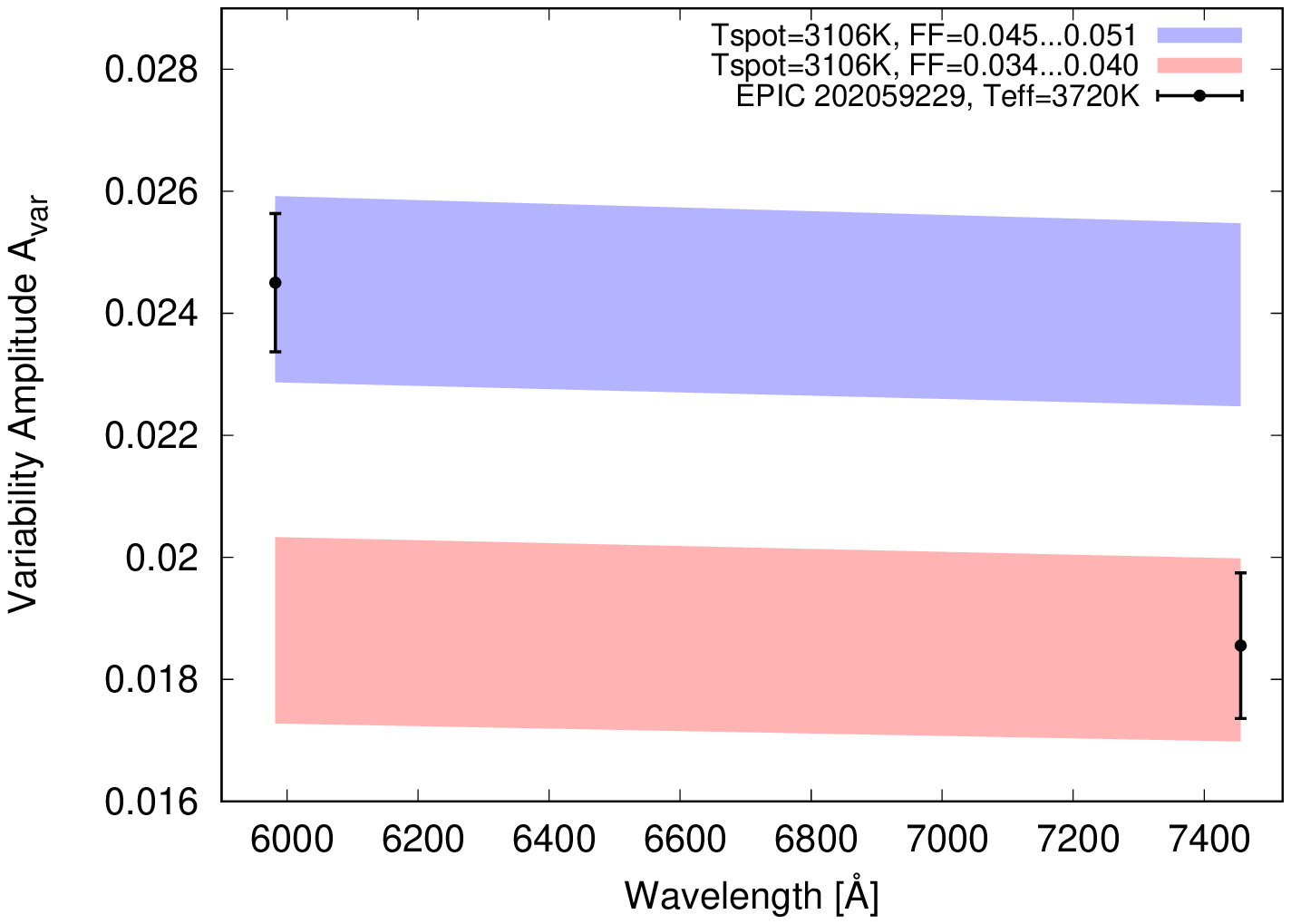}
  \caption{Photometric variability (full) amplitude over wavelength for the \textit{K2} and TESS observations of EPIC\,202059229. All models that fit either the \textit{K2} or the TESS observation are shown as the blue and red shaded areas, respectively. }
  \label{EPIC202059229_spot_Grid}
\end{figure}

\begin{table*}
\centering
\caption{ Measured rotation periods and variability amplitudes and derived spot temperatures, filling factors and spot sizes for our two targets.}
\label{amplitudes}
\begin{tabular}{lcc|cc}
\hline \hline
& \multicolumn{2}{c|}{EPIC\,202059229} & \multicolumn{2}{c}{EPIC\,245919787} \\ 
& \multicolumn{2}{c|}{TIC\,372611670} & \multicolumn{2}{c}{TIC\,434101713} \\ \hline
& \textit{K2} & TESS & \textit{K2} & TESS \\ \hline
$P_{\mathrm{rot}}$ [d] & 5.036$\pm$0.013 & 4.994$\pm$0.030 & 5.683$\pm$0.015  & 5.712$\pm$0.029 \\
adopted Amplitude $\bar{A}_{\mathrm{var}}$ & 0.0245$\pm$0.0011 & 0.0186$\pm$0.0012 & 0.0467$\pm$0.0034 & 0.0921$\pm$0.0018\\
$T_{\mathrm{spot}}$ [K] & \multicolumn{2}{c|}{3106} & \multicolumn{2}{c}{3155}  \\
model derived filling factor $f$ & 0.045 -- 0.051 & 0.034 -- 0.040 & 0.081 -- 0.095 & 0.171 -- 0.179 \\
adopted spot size $R_{\mathrm{spot}}$ [R$_{\odot}$]* & 0.3120 & 0.2780  & 0.4300 & 0.5900 \\\hline
\hline
\end{tabular}
\\
* spot size used in the light curve modeling with SOAP2.0
\end{table*}

As second step we run \begin{small}SOAP 2.0\end{small} on the light curves with $T_{\mathrm{spot}}=3106\,$K and $R_{\mathrm{spot,K2}}=0.312\,R_{\odot}$ and $R_{\mathrm{spot,TESS}}=0.278\,R_{\odot}$ as input parameters. The light curves of \textit{K2} and TESS were splitted into seven and four parts, respectively. Each part includes approximately $1.5$ rotation cycles to allow for an overlap of the models. A separate two-spot model was fitted to each part with the longitude and latitude of Spot\,1 and the longitude and latitude of Spot\,2 as free parameters. Note that for the TESS observations  Spot\,1 and Spot\,2 are not the same as for the \textit{K2} observations since spots might have disappeared and new ones appeared. We here define Spot\,1 always as the one at lower average latitude and, hence, the one that produces the deeper (primary) minimum.

\begin{figure}
  \centering
  \includegraphics[width=0.3\textwidth,angle=270]{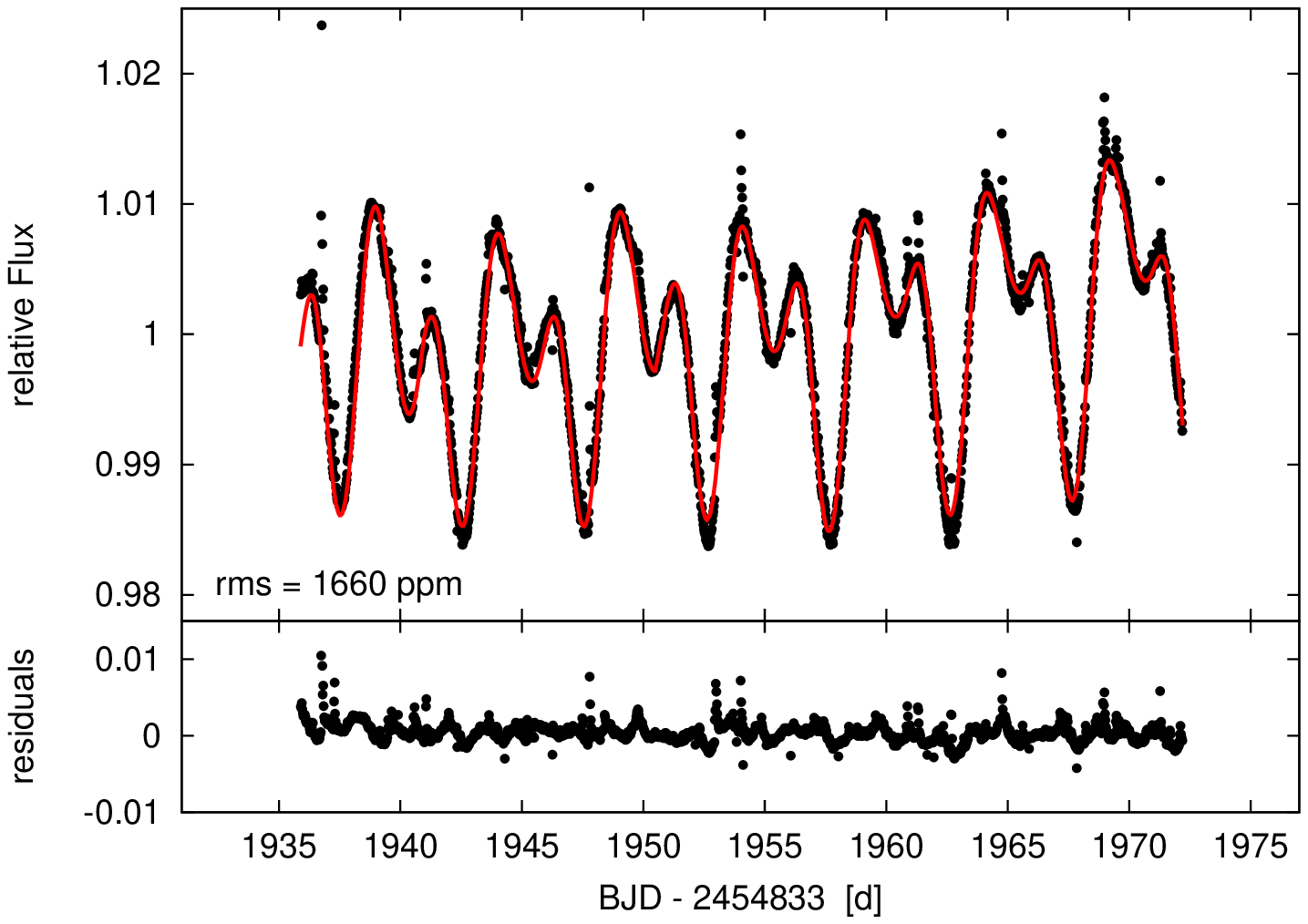}
  \includegraphics[width=0.3\textwidth,angle=270]{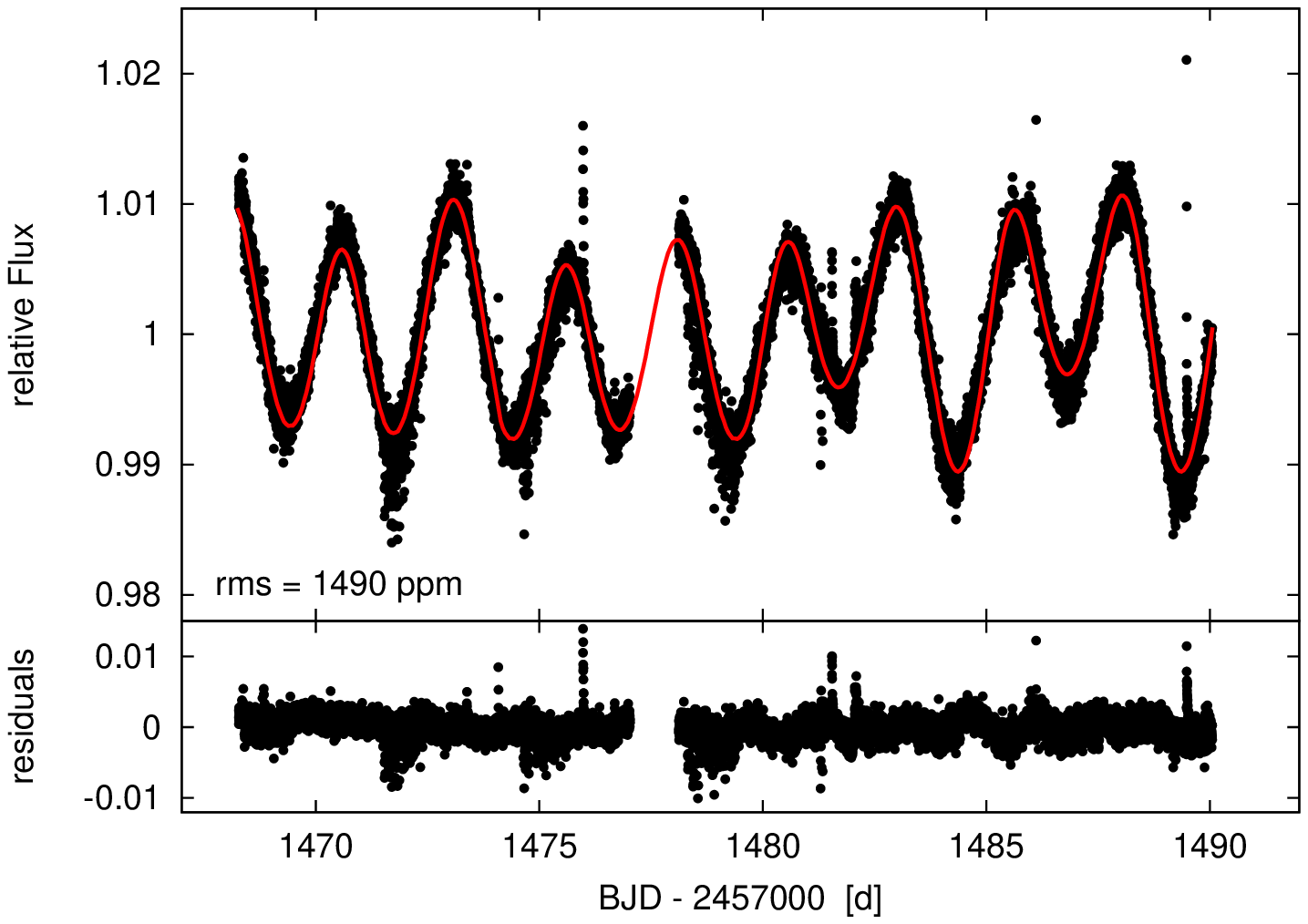}
  \caption{\textit{K2} and TESS light curves with their spot model overplotted. The light curve was splitted into several parts (see Sect.~\ref{EPIC202059229_spots}) and a separate model was fitted to each of them. The model uses two dark spots of the same size. The stellar inclination ($i=90^{\circ}$), the spot temperature (T$_{\mathrm{spot}}=3106\,$K) and the size of the spots were kept fixed for all models. \textbf{Top:} \textit{K2} light curve. A spot size of $R_{\mathrm{spot}}=0.312\,R_{\mathrm{*}}$ $(\leftrightarrow$ filling factor $f=0.049)$ was used in the model.  \textbf{Bottom:}  TESS light curve. Here a spot size of $R_{\mathrm{spot}}=0.278\,R_{\mathrm{*}}$ $(\leftrightarrow f=0.039)$ was used.}
  \label{EPIC202059229_model}
\end{figure}

Fig.~\ref{EPIC202059229_model} shows the \textit{K2} and TESS light curves together with their best fitting two-spot models. Since both spots were assumed to be of the same size (see Sect.~\ref{SOAP2}) the difference in the shape of the two light curves is caused by the different positions of the spots. The fact, that the light curves change over time means that the spots in our model are moving across the surface. Our model does not account for changes of the spot size that would indicate spot evolution. In Fig.~\ref{EPIC202059229_long_lat} the change in spot position over time is displayed. For  \textit{K2} Spot\,1 does only move marginally in latitude while Spot\,2 moves systematically to higher latitudes. For TESS the two spots that are located at similar latitudes move in opposite directions. The spots are separated in longitude by $\sim 156^{\circ}$ and $\sim 171^{\circ}$ for \textit{K2} and TESS, respectively. Since our models are based on many assumptions we do not assign error bars to our resulting spot positions. Our result should be seen as one possible solution in a degenerate problem.  

\begin{figure}
  \centering
  \includegraphics[width=0.3\textwidth,angle=270]{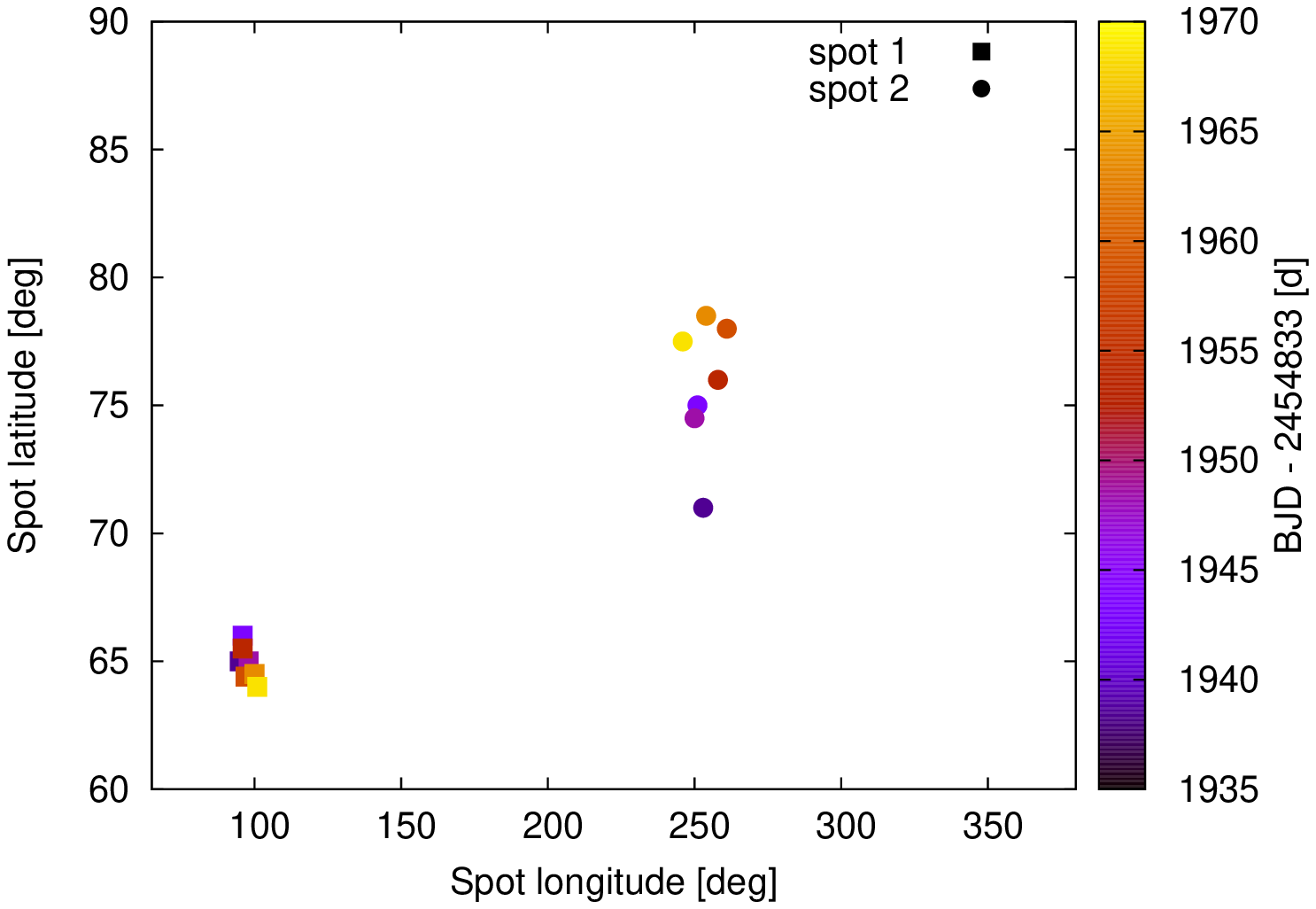}
  \includegraphics[width=0.3\textwidth,angle=270]{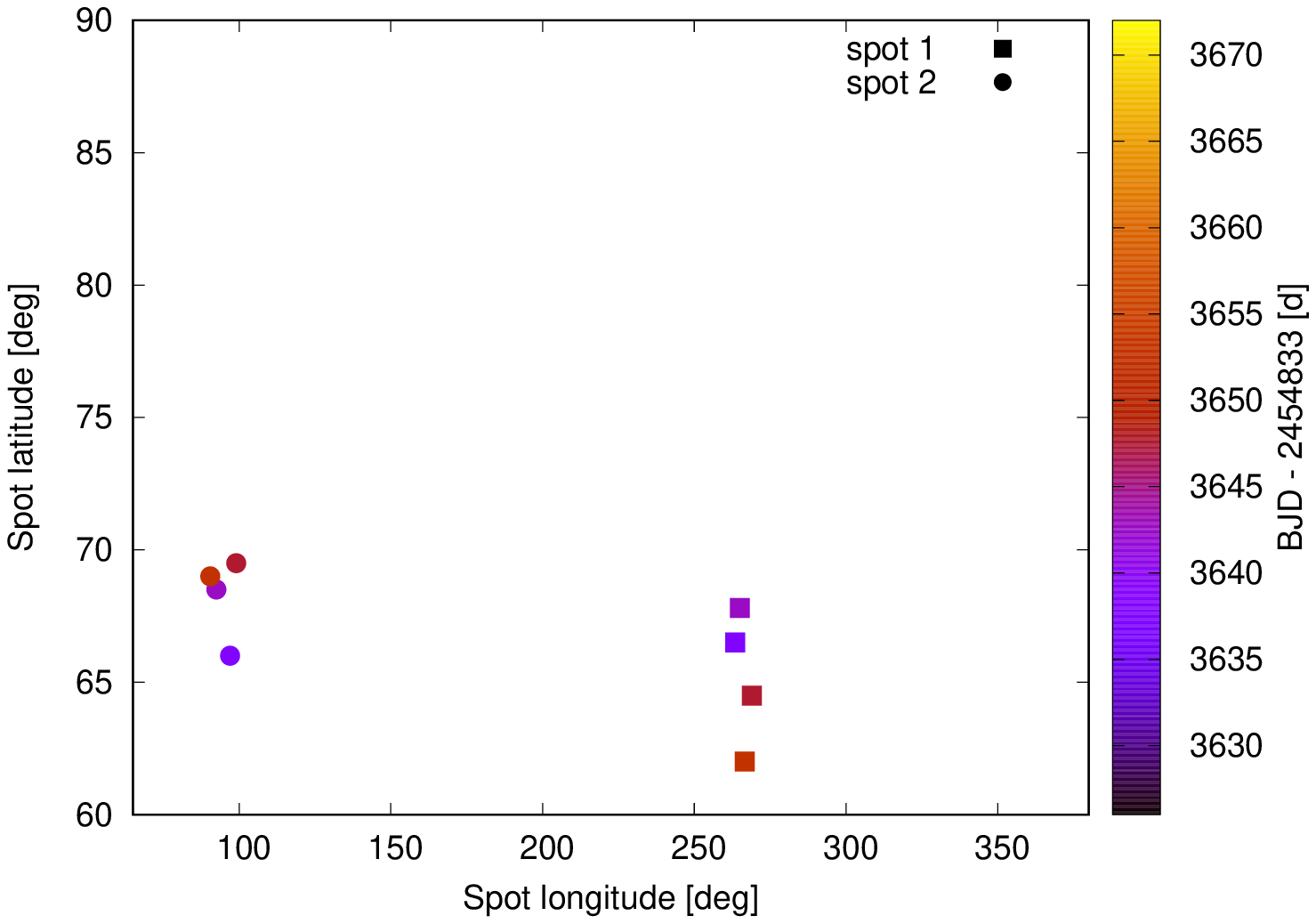}
  \caption{Change of the spot longitude and latitude of the two spots detected for EPIC\,202059229 in the \textit{K2} (top) and TESS (bottom) light curves.}
  \label{EPIC202059229_long_lat}
\end{figure}

\subsubsection{Differential rotation}
\label{DR_EPIC202059229}

\begin{figure}
  \centering
  \includegraphics[width=0.3\textwidth,angle=270]{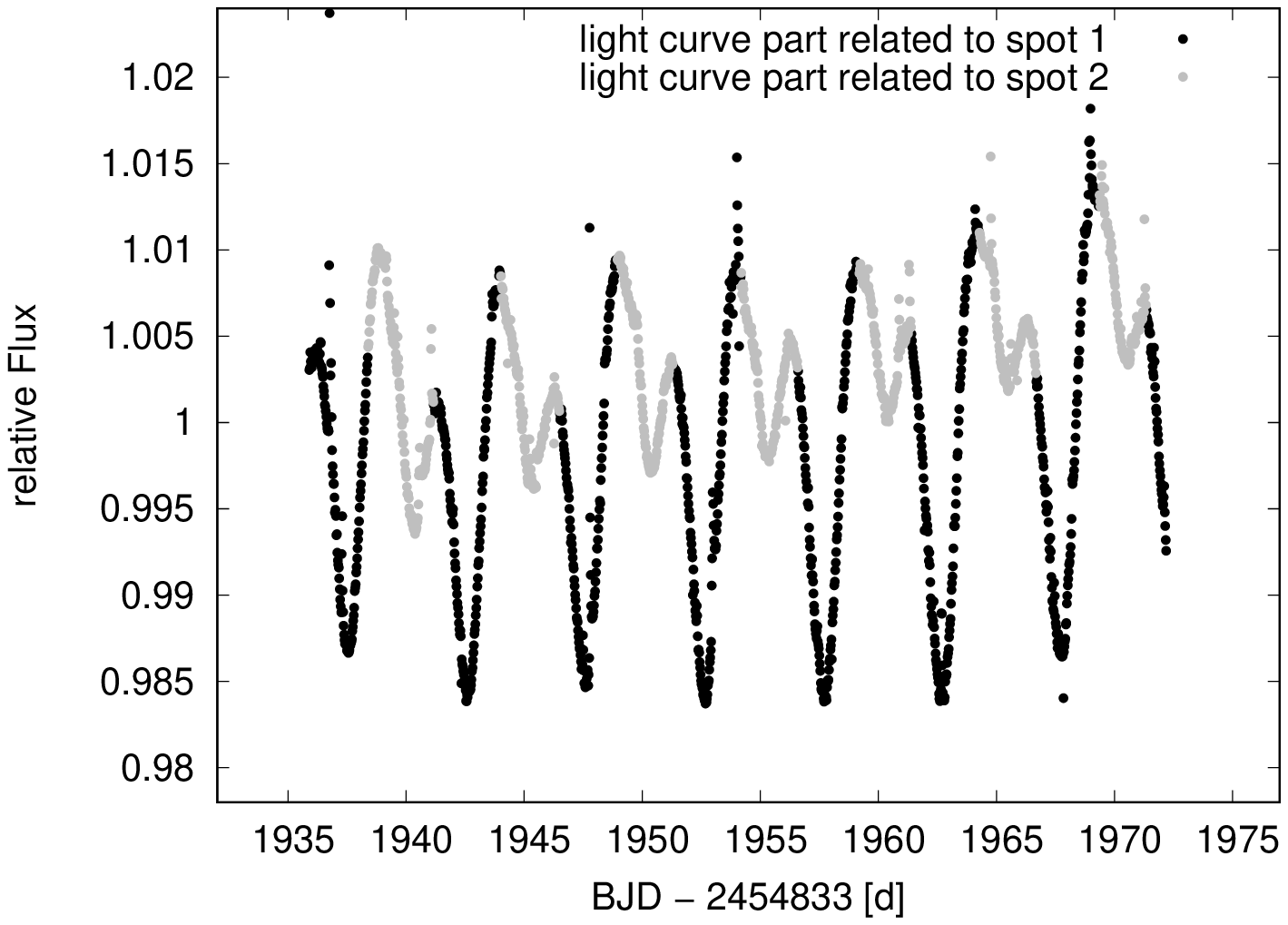}
  \includegraphics[width=0.3\textwidth,angle=270]{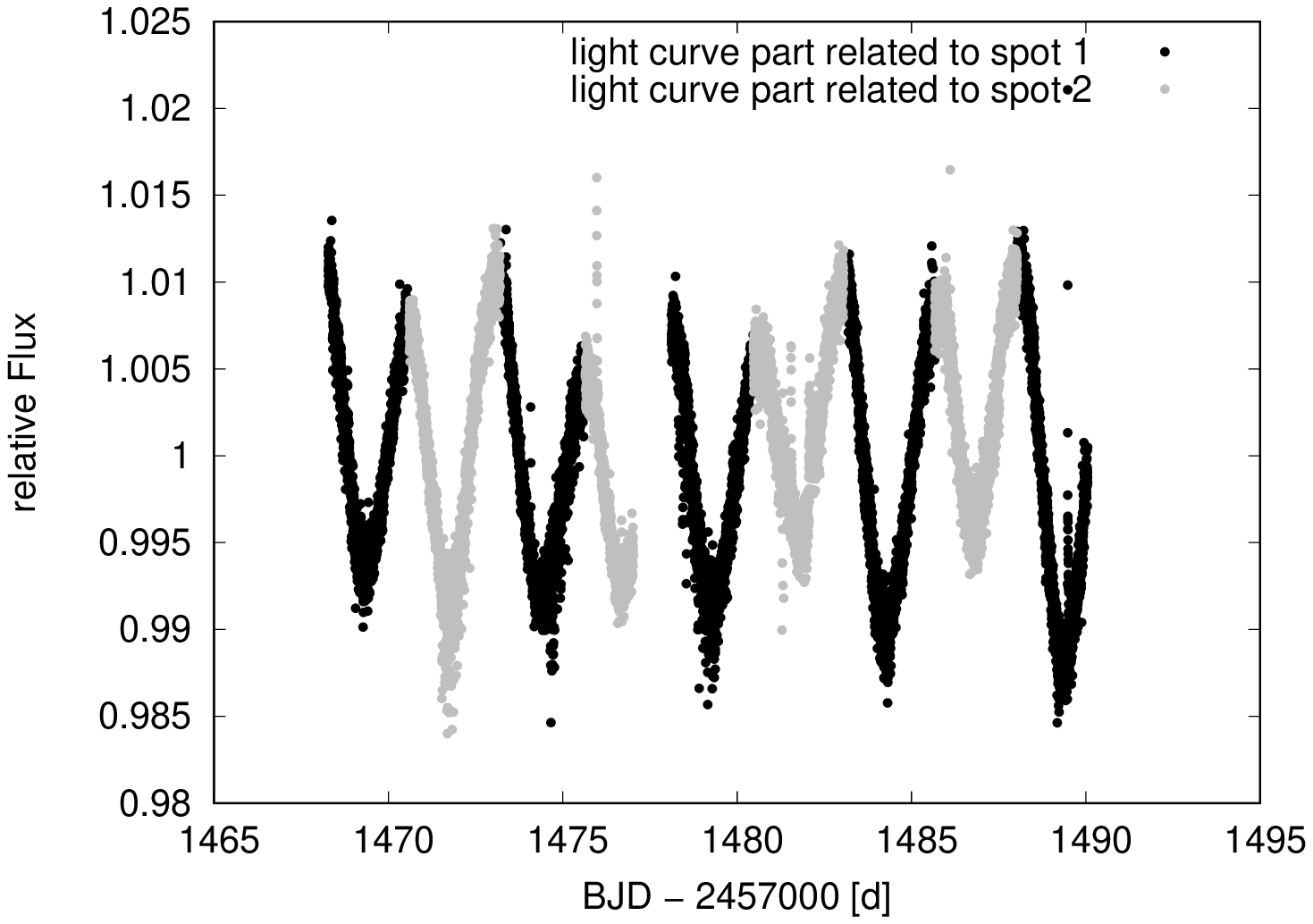}
  \caption{Light curve of EPIC\,202059229 of \textit{K2} (top) and TESS (bottom). Shown in black and in gray are the light curve intervals related to Spot\,1 and Spot\,2, respectively, that were used for the estimation of the differential rotation. The rotation period was determined separately for each interval.}
  \label{LC_EPIC202059229_parts_combined}
\end{figure}

The change of the position of the two spots with respect to each other shown in Fig.~\ref{EPIC202059229_long_lat} is a combination of spot evolution and surface differential rotation. Since the stellar dynamo is driven by differential rotation and convection \citep[e.g.][]{1966ZNatA..21.1285S} the measurement of differential rotation is essential to enhance our understanding of magnetic field generation in stars. Differential rotation has been detected in a wide sample of stars using photometric and spectroscopic methods. \citet{2013A&A...560A...4R}, for example, measured differential rotation in more than 18000 stars using light curves of the \textit{Kepler} space telescope. 

The amplitude of the differential rotation is usually measured by the rotational shear from equator to pole, $\Delta \Omega_{\mathrm{100\%}}=\Omega_{\mathrm{Eq}}-\Omega_{\mathrm{pole}}$ where $\Omega_{\mathrm{Eq}}=2\pi/P_{\mathrm{rot,Eq}}$ and $\Omega_{\mathrm{pole}}=2\pi/P_{\mathrm{rot,pole}}$ are the surface angular velocities at the equator and the pole, respectively. \citet{2013A&A...560A...4R} determined the differential rotation by the detection of a second significant period close to the dominant rotation period. We use a similar approach but instead of searching for additional peaks in the periodogram we splitted the light curves in two phase intervals: the first interval is associated to Spot\,1 (mainly all primary minima, phase $\sim0-0.5$) and the second interval is associated to Spot\,2 (mainly all secondary minima, phase $\sim0.5-1$). Fig.~\ref{LC_EPIC202059229_parts_combined} shows the two light curves of EPIC\,202059229 with a color code for the two phase intervals we defined. We then run our initial three period search methods from Sect.~\ref{period} on  each light curve interval (i.e. two intervals per observation) individually. The period that best represents each light curve intervals, $P'_{\mathrm{rot}}$, was selected by visual inspection. The uncertainties for the individual periods were determined with the formulas given by \citet{1985PASP...97..285G}. 

With the resulting values of $P'_{\mathrm{rot}}$ that are summarized in Table~\ref{Differential_Rotation} we obtained one value of $\Delta \Omega'=\Omega'_{\mathrm{spot1}}-\Omega'_{\mathrm{spot2}}$ for the \textit{K2} and one value for the TESS data. These values are, however, a lower limit to the full rotational shear (from equator to pole) and only reflect the latitudinal differences of the two spots. Therefore, we used the information on the spot positions obtained in Sect.~\ref{EPIC202059229_spots} to scale $\Delta \Omega'$ to the total rotational shear $\Delta \Omega=\Delta \Omega_{\mathrm{100\%}}$. We assumed the solar rotation profile as determined by \citet{1990ApJ...351..309S} for the scaling. For the \textit{K2} light curve the two spots are on average located at a latitude of $64.9^{\circ}$ and $75.8^{\circ}$. In the Sun the $\Delta \Omega'_{\mathrm{Sun}}$ between these two latitudes accounts for $\delta \Omega'_{\mathrm{Sun}}\sim16\%$ of the full latitudinal angular rotational velocity $\Delta \Omega_{\mathrm{Sun,100\%}}$ from equator to pole. In the TESS light curve the spots are located at a latitude of $68.3^{\circ}$ and $65.2^{\circ}$. In the Sun, these latitudes cover only $\sim5\%$ of the rotational shear. Using these percentages we were able to scale $\Delta \Omega'$ to the full equator to pole difference, $\Delta \Omega_{*,\mathrm{100\%}}$. All results of the calculations are summarized in Table~\ref{Differential_Rotation}. Within the error bars, $\Delta \Omega_{*,\mathrm{100\%}}$ is in agreement for the \textit{K2} and TESS data. Since the TESS data are less precise and, hence, the uncertainties are larger we computed a weighted average. With a final value of <$\Delta \Omega$>$=0.032 \pm 0.039$\,rad/d we could not detect a significant differential rotation in EPIC\,202059229. 

\begin{sidewaystable}
\centering
\caption{Results of the determination of the differential rotation.}
\label{Differential_Rotation}
\begin{tabular}{lr@{\,$\pm$\,}lr@{\,$\pm$\,}l|r@{\,$\pm$\,}lr@{\,$\pm$\,}l|r@{\,$\pm$\,}lr@{\,$\pm$\,}l|r@{\,$\pm$\,}lr@{\,$\pm$\,}l}
\hline \hline
& \multicolumn{8}{c|}{EPIC\,202059229}  & \multicolumn{8}{c}{EPIC\,245919787} \\ 
& \multicolumn{8}{c|}{TIC\,372611670} & \multicolumn{8}{c}{TIC\,434101713}\\ \hline
& \multicolumn{4}{c|}{\textit{K2}} & \multicolumn{4}{c|}{TESS} & \multicolumn{4}{c|}{\textit{K2}} & \multicolumn{4}{c}{TESS}\\ \hline
& \multicolumn{2}{c}{spot 1} & \multicolumn{2}{c|}{spot 2} & \multicolumn{2}{c}{spot 1} & \multicolumn{2}{c|}{spot 2} & \multicolumn{2}{c}{spot 1} & \multicolumn{2}{c|}{spot 2} & \multicolumn{2}{c}{spot 1} & \multicolumn{2}{c}{spot 2}\\ \hline
$P'_{\mathrm{rot}}$ [d] & 5.047 & 0.019 & 5.026 & 0.017 & 4.984 & 0.030 & 4.975 & 0.047 & 5.686 & 0.006 & 5.600 & 0.007 & 5.724 & 0.035 &  5.640 & 0.033 \\ 
$\Omega'$ [rad/d] & 1.245 & 0.005 & 1.250 & 0.004 & 1.261 & 0.008 & 1.263  & 0.012 & 1.105 & 0.001 & 1.122 & 0.001 & 1.098 & 0.007 & 1.114 & 0.007 \\
$\Delta \Omega'=\Omega'_{\mathrm{spot1}}-\Omega'_{\mathrm{spot2}}$ [rad/d] &  \multicolumn{4}{c|}{0.005 $\pm$ 0.006} & \multicolumn{4}{c|}{0.002$\pm$0.014}  & \multicolumn{4}{c|}{0.017$\pm$0.002} & \multicolumn{4}{c}{0.016$\pm$0.009} \\
average latitude [$^{\circ}$] & 64.9 & 0.7 & 75.8 & 2.6 & 68.3 & 1.6 & 65.2 & 2.5 & 66.6 & 1.1 & 77.4 & 3.9 & 68.6 & 0.2 & 79.8 & 0.3 \\
$\delta \Omega'_{\mathrm{Sun}}$$^{a}$ [\%] & \multicolumn{4}{c|}{16.0} & \multicolumn{4}{c|}{4.9} & \multicolumn{4}{c|}{14.4} & \multicolumn{4}{c}{13.8}  \\
$\Delta \Omega_{*,\mathrm{100\%}}$  [rad/d] &  \multicolumn{4}{c|}{0.032$\pm$ 0.039} &  \multicolumn{4}{c|}{0.051$\pm$0.291}  & \multicolumn{4}{c|}{0.119$\pm$0.012} & \multicolumn{4}{c}{0.118$\pm$0.068} \\\hline
<$\Delta \Omega$>$^{b}$  [rad/d] &\multicolumn{8}{c|}{$0.032\pm0.039$} & \multicolumn{8}{c}{$0.118\pm0.012$} \\
\hline
\end{tabular}
\\
$^{a}\delta \Omega'_{\mathrm{Sun}}$ is the percentage of the full $\Delta \Omega_{\mathrm{Sun,\mathrm{100\%}}}$ of the Sun from pole to equator the $\Delta \Omega'$ between the two different latitudes accounts for (see Sect.~\ref{DR_EPIC202059229})\\
$^{b}$<$\Delta \Omega$> weighted average of $\Delta \Omega_{\mathrm{100\%,K2}}$ and $\Delta \Omega_{\mathrm{100\%,TESS}}$
\end{sidewaystable}

\subsubsection{Flares}
\label{flares_EPIC202059229}

As shown by \citet{2020A&A...637A..22R} for the search for stellar flares the \textit{K2} short cadence light curves are better suited as long cadence light curves. Therefore, different from the period search, we used here the \textit{K2} short cadence light curves. For the \textit{K2} data we used our earlier results published in \citet{2020A&A...637A..22R} while we ran our flare detection and validation code described in Sect.~\ref{flares} on the TESS light curve. We validated 55 and 9 flare events for the \textit{K2} and TESS light curves, respectively. For each event we computed the flare energy by multiplying the equivalent duration with the quiescent luminosity. \citet{2020AJ....160...36D} showed that despite the fact that the \textit{K2} and TESS filter are very different, both instruments yield very similar flare energies. Hence, for our comparison there is no need for any conversion. We then compiled the FFD which yields information about the flare frequency as a function of flare energy. Since \citet{2020A&A...637A..22R} did not consider error bars in the power-law fit of the \textit{K2} data, we repeated the fitting process. The uncertainties for the flare frequency and the energy were calculated following the approach of \citet{2020AJ....160...36D} using the 1$\sigma$ confidence intervals for event counting statistics determined from the Poisson distribution by \citet[][their Eq. 7 and 14]{1986ApJ...303..336G} for uncertainties on the cumulative flare rates and Eq. 6 from \citet{2006AN....327..862V} for uncertainties on the equivalent duration of the flares.

It is well known that the completeness of the FFD towards lower energies depends on the noise level of the light curves. This completeness limit is seen as a break in the power-law for low energies in the FFDs. We computed this energy above which we are able to detect all flares, $E_{\mathrm{min}}$, using the method described by \citet{2020A&A...637A..22R} and \citet{2022A&A...665A..30S}. It consists in estimating the minimum energy that a flare would need to have to be detected at our detection threshold of 3$\sigma$ using the flare template of \citet{2014ApJ...797..122D} and an assumed flare duration of 360\,s. In this simulation we varied the amplitude of the flare template until five and three data points (that correspond to $\sim$240\,s flare duration) for \textit{K2} and TESS, respectively,  were above the 3$\sigma$ threshold. Hence, the 360\,s assumed total flare duration allows for points of the flare to be below the detection threshold. The resulting equivalent duration was then multiplied by the quiescent luminosity to obtain $E_{\mathrm{min}}$. We found our completeness limits to be log$(E_{\mathrm{min,K2}})[\mathrm{erg}]=31.78$ and log$(E_{\mathrm{min,TESS}})[\mathrm{erg}]=32.25$. We fitted the FFDs of \textit{K2} and TESS with a power-law taking the error bars into account. Only flare events above the calculated completeness limit were considered in the fit.

Fig.~\ref{FFD_EPIC202059229} shows the FFDs obtained with \textit{K2} and TESS together with the best-fitting models. The resulting power-law fits are in excellent agreement with each other ($\beta_{\mathrm{K2}}=-0.77\pm0.10$, $\beta_{\mathrm{TESS}}=-0.76\pm1.05$) and with the fit of \citet[][$\beta=-0.77\pm0.44$]{2020A&A...637A..22R} for the \textit{K2} data within the uncertainties. In Fig.~\ref{FFD_EPIC202059229} we also show the FFD presented by \citet[][extracted from their Fig. 5]{2015MNRAS.449.3015R}. Considering their reported uncertainty of $\sim40\%$ in the bolometric luminosities, and, hence their large error bars on the flare energies, their slope is consistent with our findings.

The comparison of the total flare rate for different instruments is not straight forward since flares occur with a wide range of energies and each observation is characterized by its own completeness limit. Hence, we computed a specific flare rate \citep{2019ApJ...871..241D} where we include all flares with energies above a predefined energy threshold. We calculated the specific flare rate $\nu_{32.5}$, with the subscript denoting the log energy that rate is evaluated at. We chose the value of log$(E)[\mathrm{erg}]=32.5$ because it is well above the completeness limits for \textit{K2} and TESS. The specific flare rates found for EPIC\,202059229 are $\nu_{32.5,\mathrm{K2}}=0.23\,\mathrm{d}^{-1}$ and $\nu_{32.5,\mathrm{TESS}}=0.19\,\mathrm{d}^{-1}$. Considering the uncertainties on the determination of the flare rates the specific flare rate is in excellent agreement for the two light curves.

\begin{figure}
  \centering
  \includegraphics[width=0.35\textwidth,angle=270]{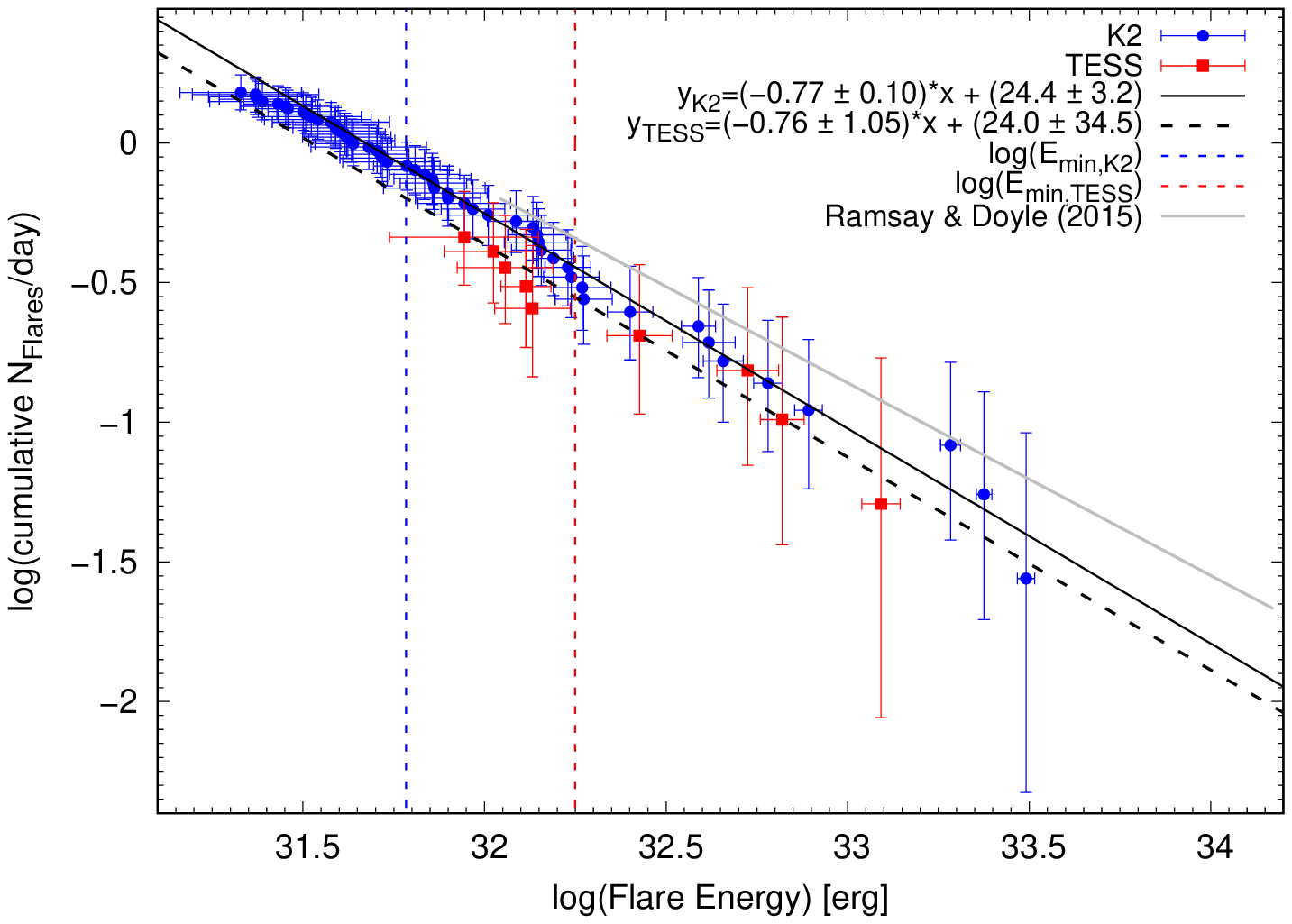}
  \caption{Cumulative flare frequency distribution for the \textit{K2} and TESS data of EPIC\,202059229. The completeness limits are marked as blue and red dashed vertical lines. The best-fitting power-law is given as the black solid (\textit{K2}) and black dashed (TESS) lines. The results of \citet{2015MNRAS.449.3015R} from the same \textit{K2} data is shown as the grey solid line.}
  \label{FFD_EPIC202059229}
\end{figure}

\subsection{EPIC\,245919787 / TIC\,434101713}
\label{EPIC245919787_results}

The M1.0V dwarf EPIC\,245919787 (TIC\,434101713, BD-13\,6424, TYC\,5832-666-1) is a member of the $\beta$ Pic moving group \citep{2009AJ....137.3632L,2017A&A...607A...3M} and, hence, has a young age of $\sim$24\,Myrs \citep{2017AJ....154...69S}. Studies of the spectral energy distribution by \citet{2017MNRAS.469..579B} did not show evidence for an infrared excess, and they concluded that there is no disk around EPIC\,245919787. \textit{ROSAT} observations identified EPIC\,245919787 as an X-ray source \citep{1998A&A...335..467T}. Lucky imaging and high resolution imaging studies by \citet{2010A&A...520A..54B}, \citet{2012A&A...539A..72D}, \citet{2013ApJ...777..160B}, \citet{2014A&A...568A..26E,2015A&A...580A..88E,2016A&A...590A..13E} and \citet{2016A&A...594A..63G} did not find any companions. 

Using ASAS photometry \citet{2010A&A...520A..15M} measured a rotation period of  $P_{\mathrm{rot}}=5.68 \pm 0.05$\,d. \citet{2016A&A...591A..43D} reported a lower limit for the differential rotation of EPIC\,245919787 which they obtained by analyzing data from ASAS. Their value of $\Delta \Omega=\Omega_{\mathrm{Eq}}-\Omega_{\mathrm{pole}}= 0.031$\,rad/d, indicates that the equator rotates faster than the poles (solar-like differential rotation). Furthermore, \citet{2017A&A...606A..58D} found an activity cycle with a period of $P_{\mathrm{cyc}}=1695\pm 259$\,d in the observations of ASAS.

EPIC\,245919787 was observed with \textit{K2} in campaign C12 for a duration of $\sim$79\,d in short and long cadence mode. The detailed rotation and activity analysis of the \textit{K2} data can be found in \citet{2020A&A...637A..22R}. Approximately two years later EPIC\,245919787  was observed in the TESS prime mission in sector 2 for $\sim$25\,d with a 2-min cadence. The TESS data were previously analyzed by \citet{2019ApJ...881....9H,2020ApJ...895..140H} and \citet{2020AJ....159...60G}. The former refined the rotation period to $P_{\mathrm{rot}}=5.6820\pm0.0004$\,d while the latter analyzed the flares and the FFD in detail.

In the following sections we present the results of the analysis of the \textit{K2} and TESS data with the comparison of the stellar activity. All stellar parameters of EPIC\,245919787 used in our analysis are summarized in Table~\ref{stellar_param}.

\subsubsection{Rotation period and variability amplitude}

For the determination of the rotation period we used the EVEREST and the PDCSAP light curves for \textit{K2} and TESS, respectively. We also inspected the TESS SAP light curve. Since the pipeline detrending worked fine for EPIC\,245919787 we decided to use the PDCSAP light curve directly. After removing the data points with the chosen quality flags (see Sect.~\ref{EPIC202059229_rotation}) from the TESS light curve both light curves were normalized by dividing all data points by the median flux. On these final light curves we applied the period search methods described in Sect.~\ref{period}. The phase-folded light curves with our final refined period are shown in Fig.~\ref{best_Phasfolded_LC_0245919787}. The results of the initial period search can be found in Appendix~\ref{period_plots} in Fig.~\ref{Phasfolded_LC_0245919787_K2} and Fig.~\ref{Phasfolded_LC_0245919787_TESS}. We found values of the rotation period of $P_{\mathrm{rot,K2}}=5.683\pm0.015$ and $P_{\mathrm{rot,TESS}}=5.712\pm0.029$ for \textit{K2} and TESS, respectively. The period determined from the TESS light curve is, thus, in agreement with the \textit{K2} measurement.

\begin{figure}
  \centering
  \includegraphics[width=0.35\textwidth,angle=270]{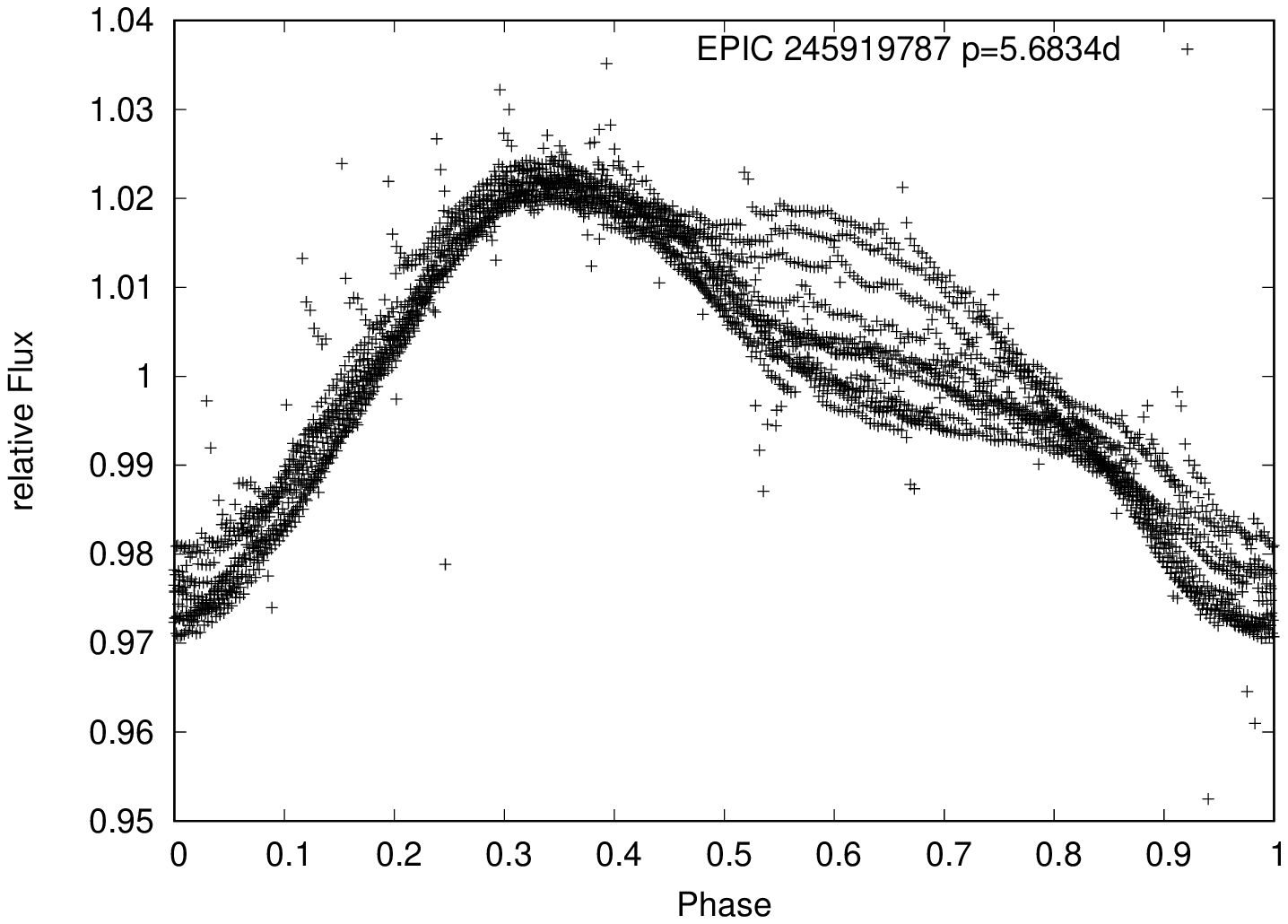}
  \includegraphics[width=0.35\textwidth,angle=270]{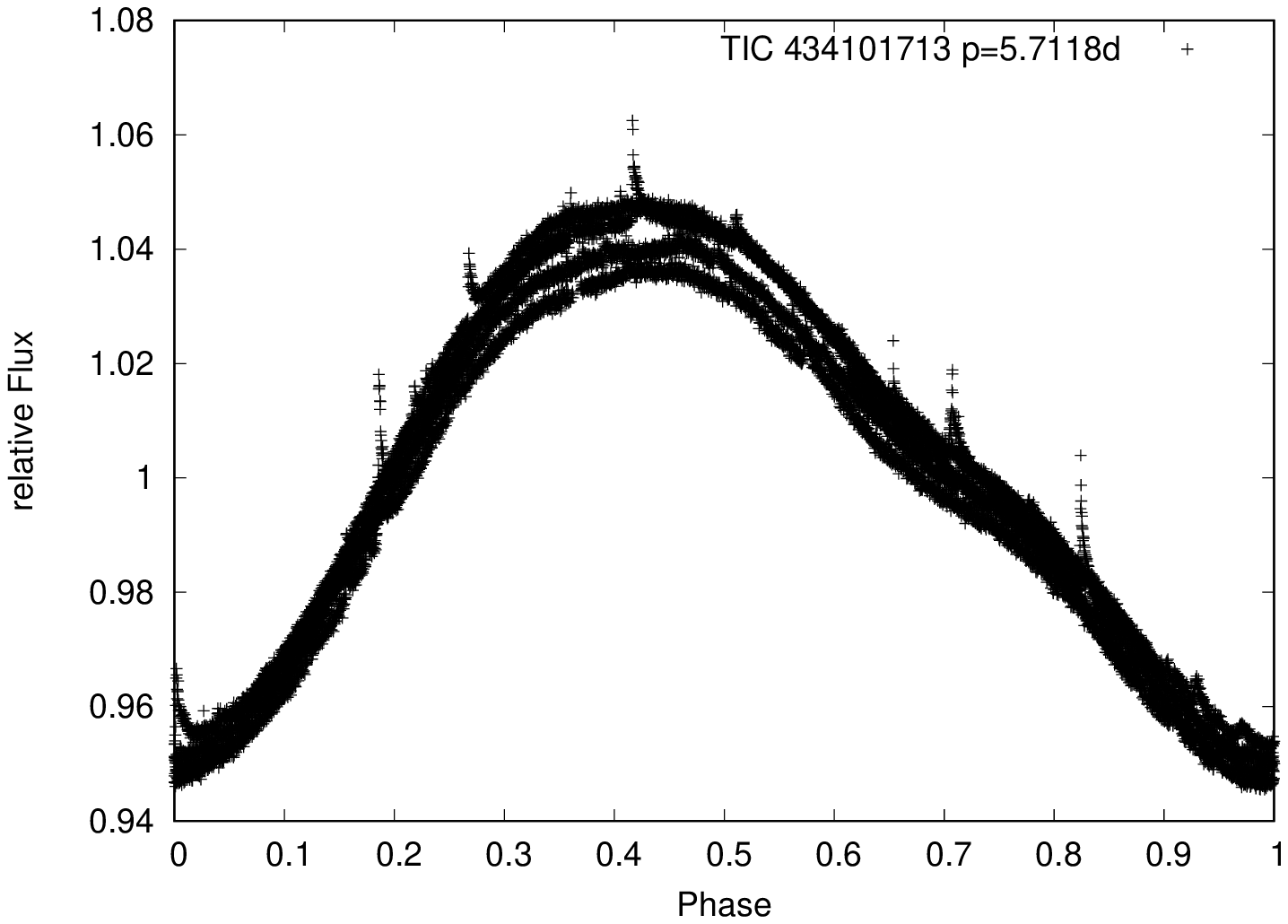}  
  \caption{Light curves of EPIC\,245919787 folded with the refined rotation period (see Sect.~\ref{period}). \textbf{Top:} Light curve observed with \textit{K2} in campaign C12  \textbf{Bottom:} Light curve observed with TESS in sector 2.}
  \label{best_Phasfolded_LC_0245919787}
\end{figure}

The light curves show that the variability amplitude is changing with time. Moreover, the amplitude is larger in the TESS light curve compared to the \textit{K2} light curve. To quantify the change in amplitude, we determined $A_{\mathrm{var}}$ individually for each rotation. In contrast to EPIC\,202059229 where the primary minimum can be distinguished from the secondary one, EPIC\,245919787 shows only a second hump in its light curve. We used this fact and did not only measure the amplitude from the maximum to the primary minimum but also from the primary minimum to the following maximum. In total, we measured 25 and 8 values for $A_{\mathrm{var}}$ for \textit{K2} and TESS, respectively.  We calculated  $\bar{A}_{\mathrm{var}}$ by taking the average of all individual amplitudes. Fig.~\ref{EPIC245919787_time_ampl} shows the variation of $A_{\mathrm{var}}$. The mean value in the TESS light curve, $\bar {A}_{\mathrm{var,TESS}}=0.0921 \pm 0.0018$, is approximately 2 times larger than the mean \textit{K2} value, $\bar{A}_{\mathrm{var,K2}}=0.0467 \pm 0.0034$. This amplitude change is significant at $\sim 13\,\sigma$.

\begin{figure}
  \centering
  \includegraphics[width=0.35\textwidth,angle=270]{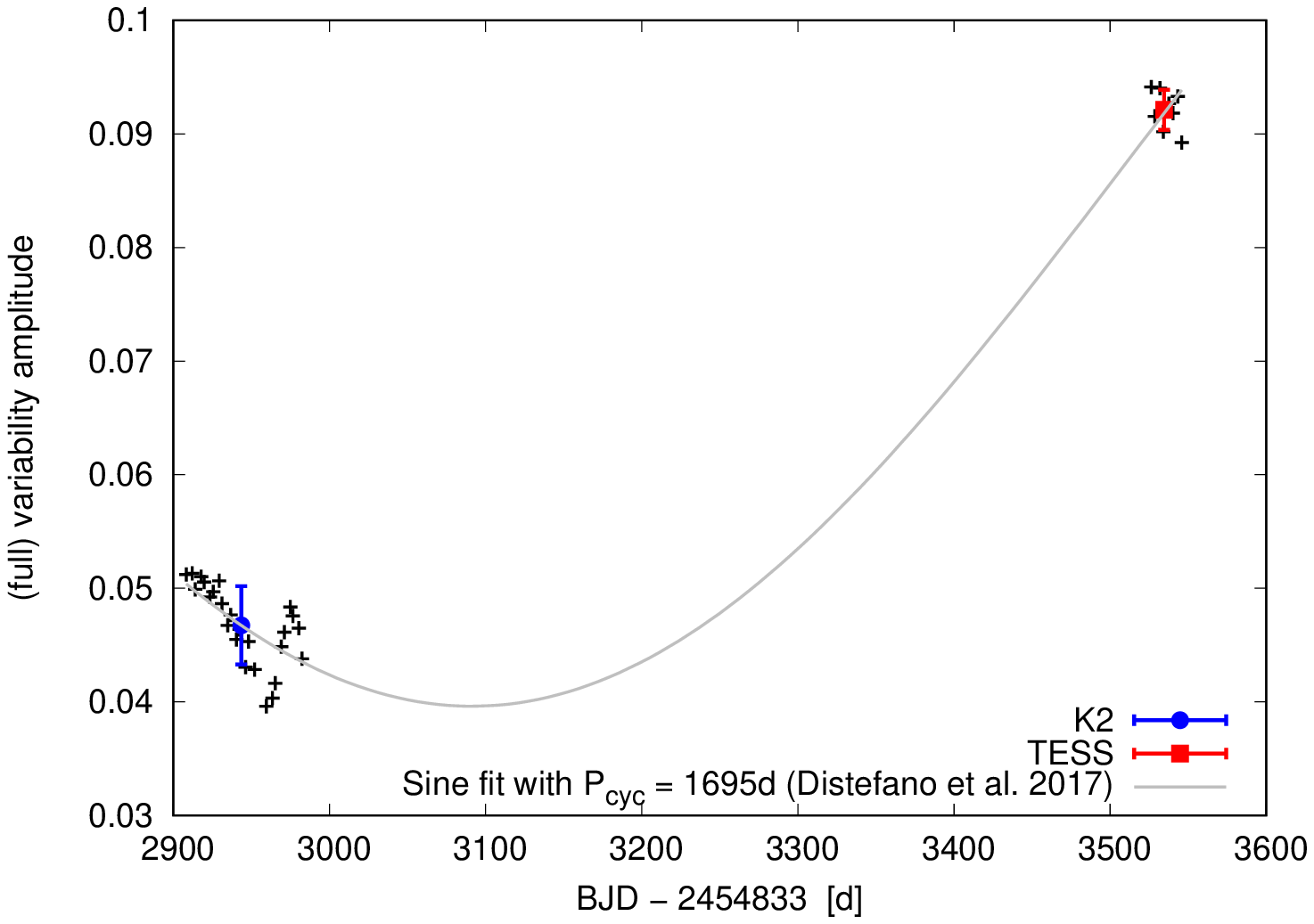}
  \caption{Variability amplitude for the individual rotation cycles in the \textit{K2} and TESS light curves of EPIC\,245919787. The blue circle and the red square mark the $\bar{A}_{\mathrm{var}}$ values that we then used in the spot modeling. The best fitting sine function with the cycle period found by \citet{2017A&A...606A..58D} in ASAS data is also shown.}
  \label{EPIC245919787_time_ampl}
\end{figure}

\citet{2017A&A...606A..58D} analyzed long-term photometric time series collected by the ASAS survey and found a cycle period of $P_{\mathrm{cyc}}=1695\pm295$\,d for EPIC\,245919787. Since \citet{2017A&A...606A..58D} do not give the phase and amplitude of their detected activity cycle we applied a simple sine fit to our data but keeping the cycle period fixed. The best fitting sine function with this cycle period is also shown in Fig.~\ref{EPIC245919787_time_ampl}. The change of $A_{\mathrm{var}}$ in the $\sim 2$ years time span between the {\it K2} and TESS observations is in agreement with the cycle presented by \citet{2017A&A...606A..58D}. However, the short observational baseline between the \textit{K2} and TESS light curves is not sufficient to perform an independent analysis of the cycle.

\begin{figure*}
  \centering
  \includegraphics[width=0.35\textwidth,angle=270]{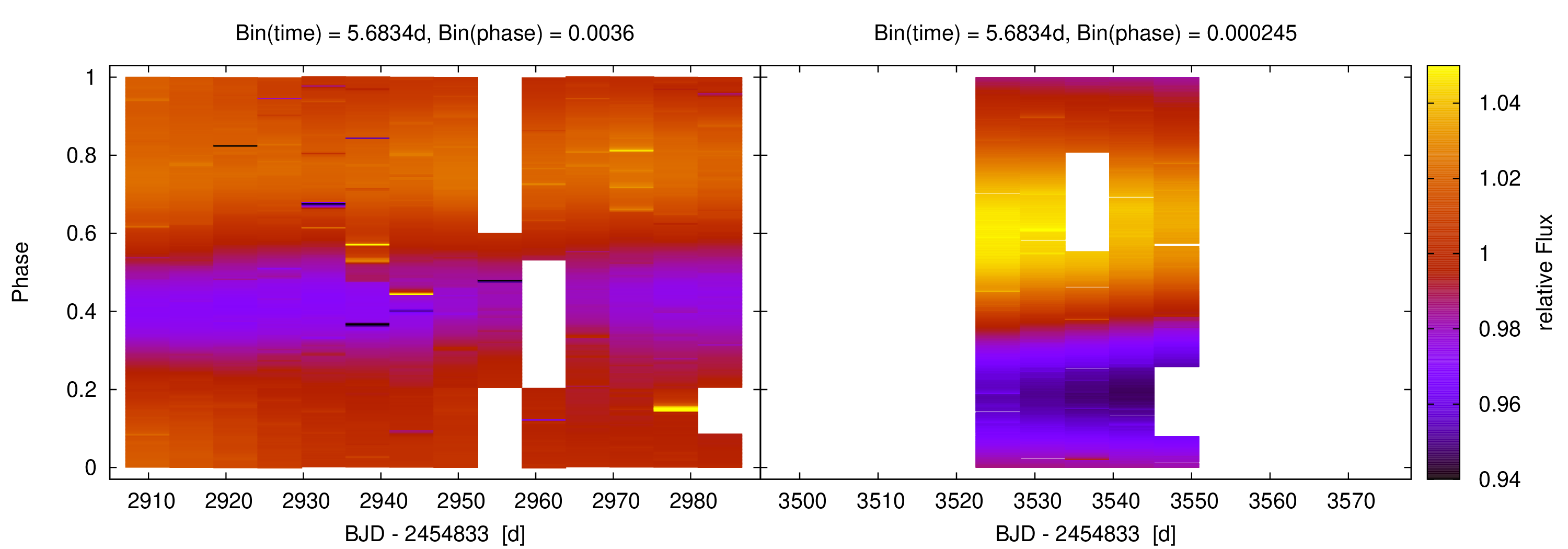}
  \caption{Same as Fig.~\ref{EPIC202059229_time_phase_flux} but for EPIC\,245919787. \textbf{Left:} Observations of \textit{K2}. The light curve was splitted into the $\sim 14$ rotation cycles present in the total light curve. \textbf{Right:} Observations of TESS. The TESS data only consists of five rotation cycles. Both panels were plotted in the same time range and the same color bar to show the light curves in the same scale. A phase shift of the TESS observations and an increased variability amplitude is clearly visible.}
  \label{EPIC245919787_time_phase_flux}
\end{figure*}

In Fig.~\ref{EPIC245919787_time_phase_flux} the 3D representation of the light curves is shown (light curve in terms of phase over time with the relative flux as color code). Although the change in $P_{\mathrm{rot}}$ is not significant, there is a change in the rotation phase by $\sim$0.2 and in $A_{\mathrm{var}}$. In contrast to EPIC\,202059229, Fig.~\ref{EPIC245919787_time_phase_flux} reveals that the level of stellar activity has changed within the observing baseline of $\sim 2$ years.

\subsubsection{Light curve modeling}
\label{EPIC245919787_spots}

Analogous to Sect.~\ref{EPIC202059229_spots} we compared the values of $\bar{A}_{\mathrm{var}}$ that we measured in the \textit{K2} and TESS light curves to the grid of variability amplitudes. For EPIC\,245919787 we fixed the spot temperature for the simulation to $T_{\mathrm{spot}}=3155$\,K, the value we calculated using Eq.~4 in \citet{2019ApJ...876...58N} with $T_{\mathrm{eff}}=3821$\,K (see Table~\ref{stellar_param}). With the method described in Sect.~\ref{amplitude_sim} we computed a final grid that contains 64 individual models with $f$ ranging from $0.075 - 0.202$. From these 64 models, 15 and nine for \textit{K2} and TESS, respectively, match the measured amplitudes. The range of models that fit either the \textit{K2} or the TESS observation is shown in Fig.~\ref{EPIC245919787_spot_Grid}. The corresponding filling factors $f_{\mathrm{K2}}=0.081 - 0.095$ for \textit{K2} and  $f_{\mathrm{TESS}}=0.171 - 0.179$ for TESS translate into $R_{\mathrm{spot,K2}}= (0.40 - 0.44)\,R_{\mathrm{*}}$ and  $R_{\mathrm{spot,TESS}}= (0.58 - 0.60)\,R_{\mathrm{*}}$. The rotation periods, amplitudes and the derived spot parameters are summarized in Table~\ref{amplitudes}.

\begin{figure}
  \centering
  \includegraphics[width=0.35\textwidth,angle=270]{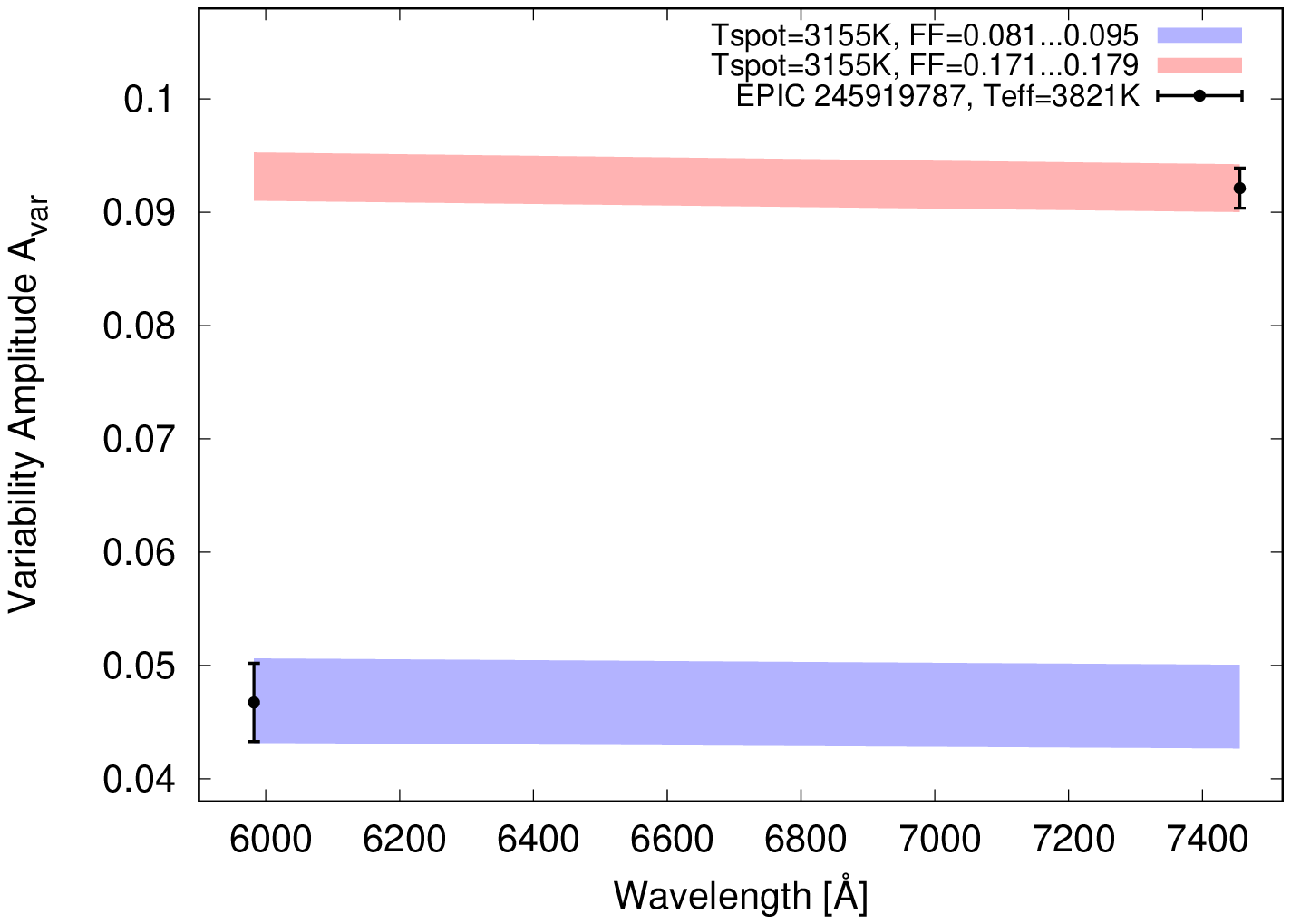}
  \caption{Photometric variability (full) amplitude over wavelength for the \textit{K2} and TESS observations of EPIC\,245919787. All models that fit either the \textit{K2} or the TESS observations are shown as the blue and red shaded areas, respectively.}
  \label{EPIC245919787_spot_Grid}
\end{figure}

The spot temperature $T_{\mathrm{spot}}=3155$\,K and the spot sizes $R_{\mathrm{spot}}=0.43\,R_{\mathrm{*}}$ for \textit{K2} and $R_{\mathrm{spot}}=0.59\,R_{\mathrm{*}}$ for TESS were then used as input for the spot modeling with \begin{small}SOAP 2.0\end{small}. To account for the motion of the spots on the stellar surface we splitted the light curves in ten and five parts for \textit{K2} and TESS, respectively, and a separate model was fitted to each part. The resulting light curves with the best fitting two-spot-model overplotted is shown in Fig.~\ref{EPIC245919787_model}. 

\begin{figure}
  \centering
  \includegraphics[width=0.3\textwidth,angle=270]{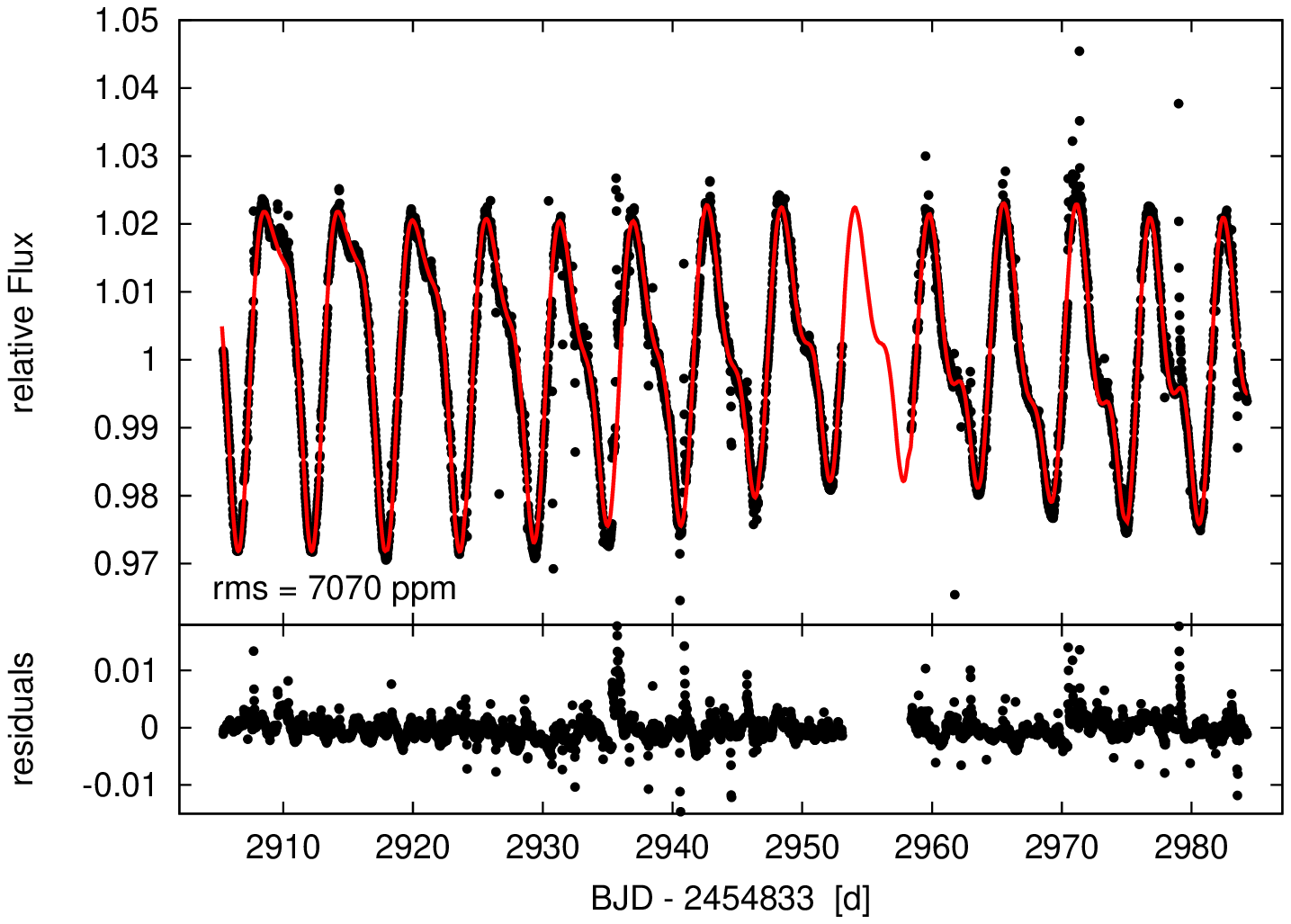}
  \includegraphics[width=0.3\textwidth,angle=270]{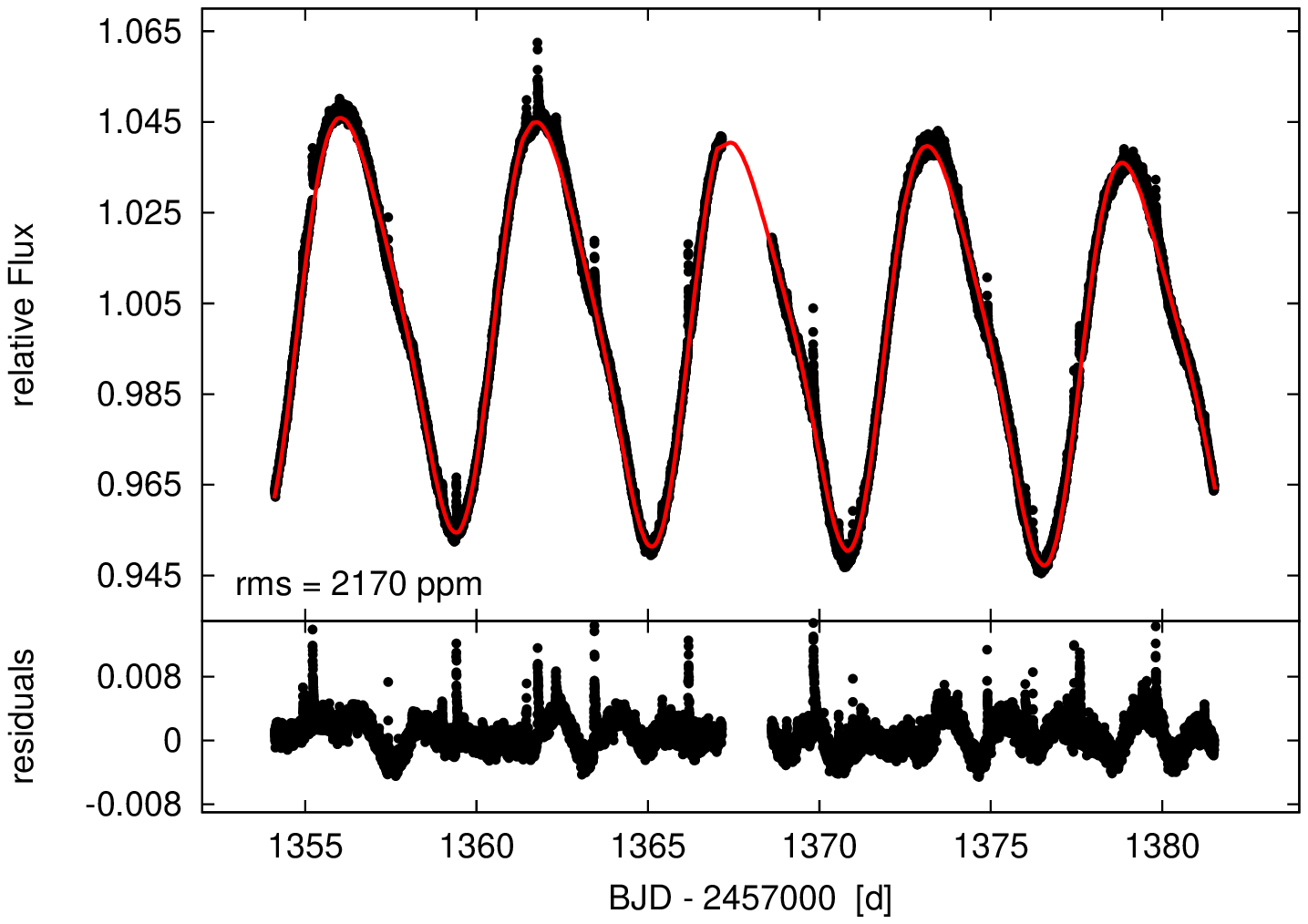}
  \caption{\textit{K2} and TESS light curves of EPIC\,245919787 with their spot model overplotted. The stellar inclination ($i=90^{\circ}$), the spot temperature (T$_{\mathrm{spot}}=3155\,$K) and the size of the spots were kept fixed for all models. \textbf{Top:} \textit{K2} light curve. A spot size of $R_{\mathrm{spot}}=0.43\,R_{\mathrm{*}}$ $(\leftrightarrow f=0.09)$ was used in the model.  \textbf{Bottom:}  TESS light curve. Here a spot size of $R_{\mathrm{spot}}=0.59\,R_{\mathrm{*}}$ $(\leftrightarrow f=0.17)$ was used.}
  \label{EPIC245919787_model}
\end{figure}

Our spot modeling allows us to monitor the spot position over time. The result for the \textit{K2} and TESS observations is given in Fig.~\ref{EPIC245919787_long_lat}. 
In the TESS light curve there is only a slight motion of both spots. In the \textit{K2} data Spot\,1 is also moving marginally while the position of Spot\,2 increases by $\sim 10^{\circ}$ in latitude. The two spots are separated in longitude by $\sim 230^{\circ}$ and $\sim 109^{\circ}$ for \textit{K2} and TESS, respectively.

\begin{figure}
  \centering
  \includegraphics[width=0.3\textwidth,angle=270]{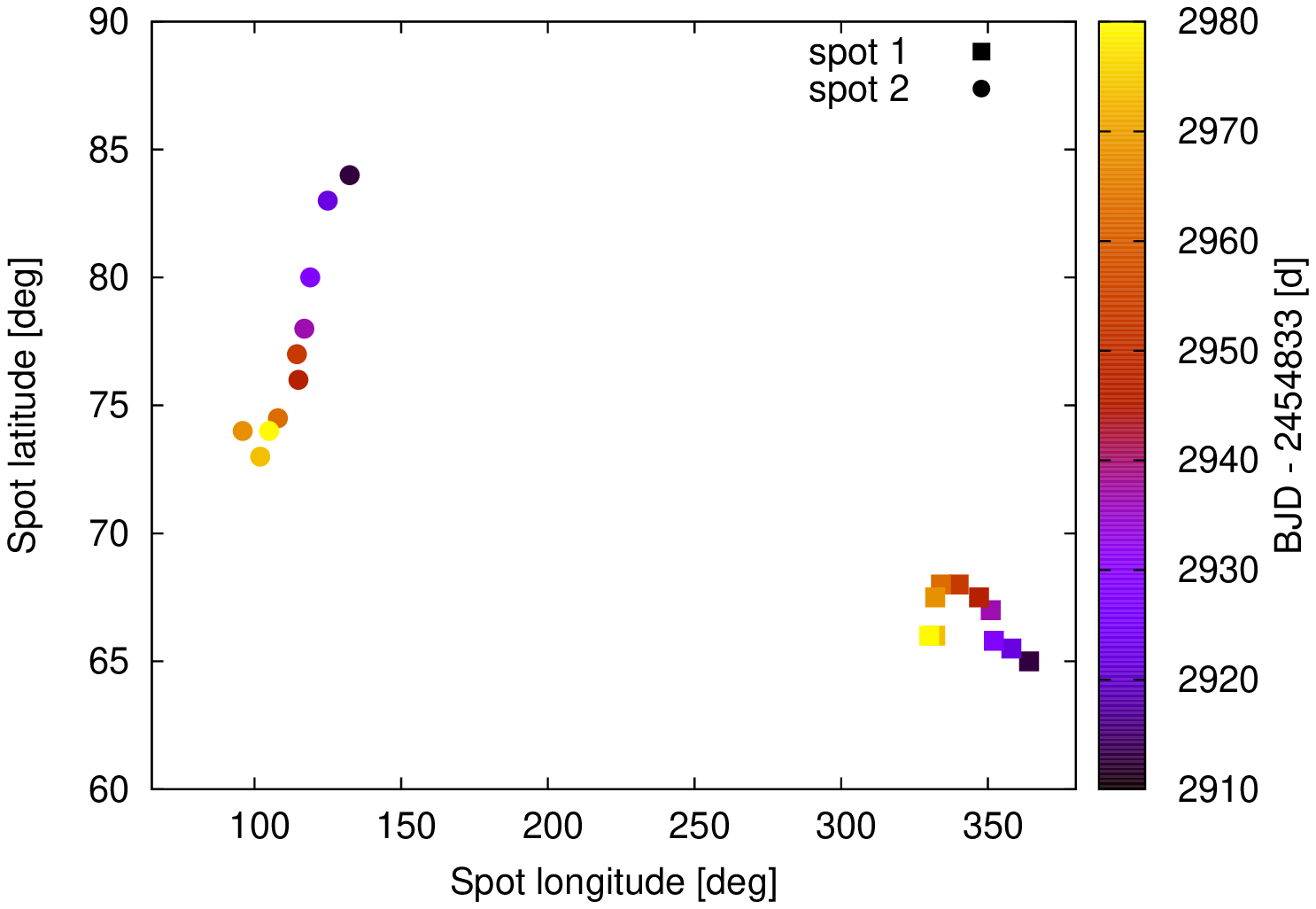}
  \includegraphics[width=0.3\textwidth,angle=270]{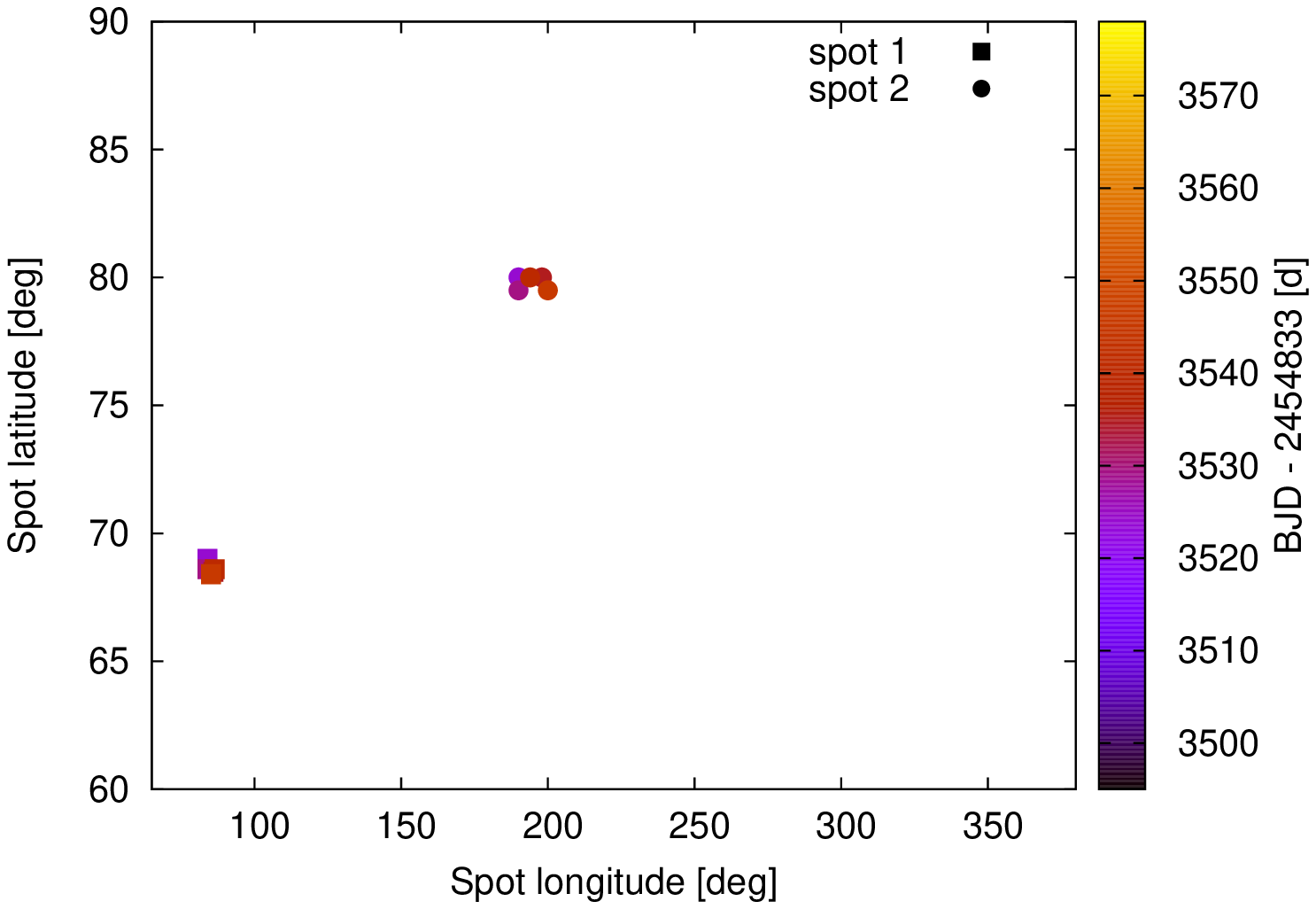}
  \caption{Change of the spot longitude and latitude of the two spots detected in the light curve of EPIC\,245919787 for the \textit{K2} (top) and TESS (bottom) light curves.}
  \label{EPIC245919787_long_lat}
\end{figure}

\subsubsection{Differential rotation}

\begin{figure}
  \centering
  \includegraphics[width=0.3\textwidth,angle=270]{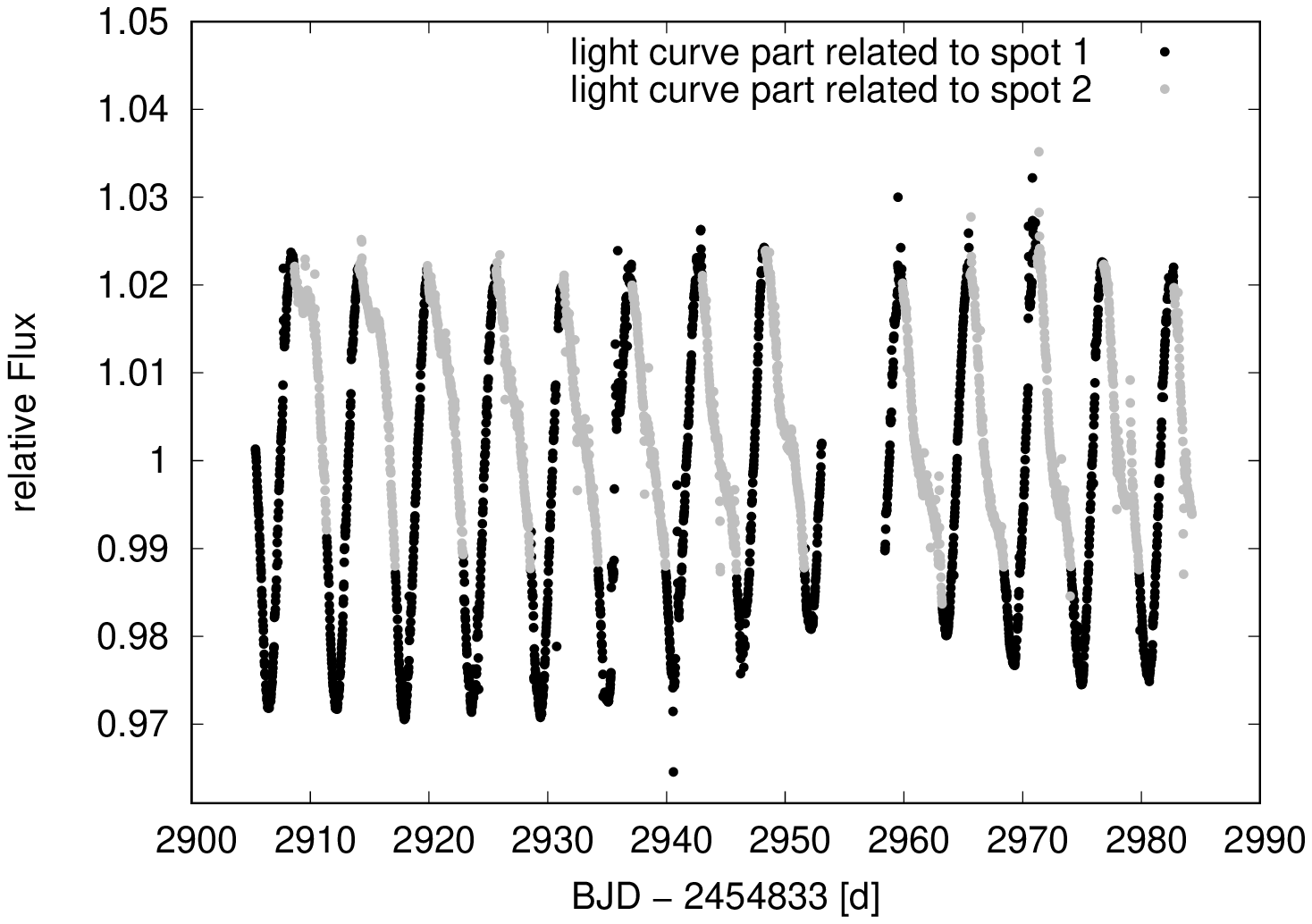}
  \includegraphics[width=0.3\textwidth,angle=270]{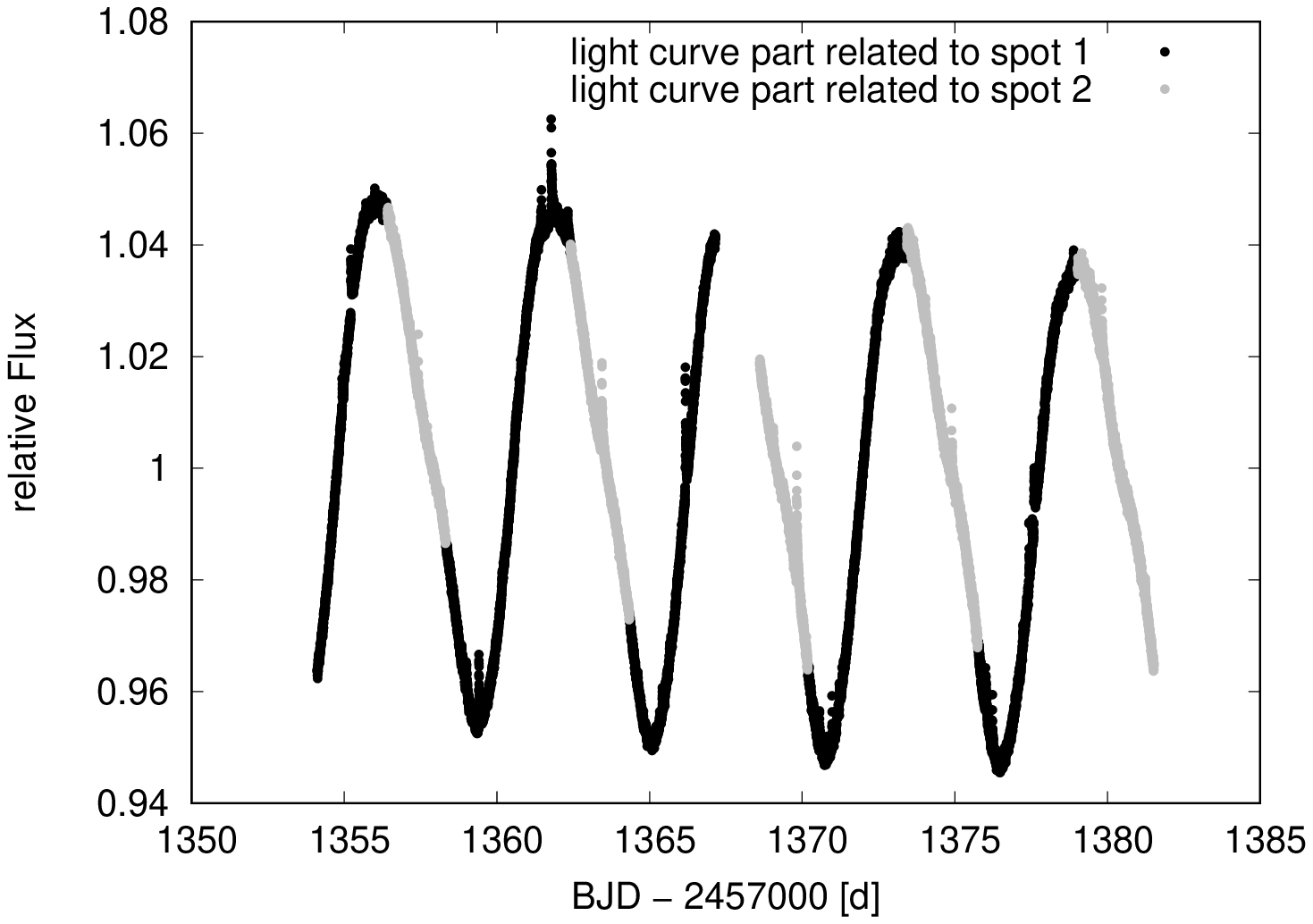}
  \caption{Same as Fig.~\ref{LC_EPIC202059229_parts_combined} but for EPIC\,245919787. The  \textit{K2} and TESS light curves are shown in the top and bottom panel, respectively. The black and gray circles denote the phase intervals related to Spot 1 and Spot 2 that were used for the estimation of the differential rotation, respectively.}
  \label{LC_EPIC245919787_parts_combined}
\end{figure}

A lower limit for the differential rotation in EPIC\,245919787 was reported by \citet{2016A&A...591A..43D}. Although we do not have such a long observational baseline than the one provided by the ASAS data, we were able to estimate the differential rotation with the method described in Sect.~\ref{DR_EPIC202059229}. The splitting of the light curves in a phase interval for Spot\,1 (mainly the primary minimum, phase $\sim -0.1-0.4$) and in a phase interval for Spot\,2 (mainly the second hump, phase $\sim 0.4-0.9$) that is shown in Fig.~\ref{LC_EPIC245919787_parts_combined} resulted in different best-fitting periods $P'_{\mathrm{rot}}$. The values of $\Delta \Omega'=\Omega'_{\mathrm{spot1}}-\Omega'_{\mathrm{spot2}}$ for the \textit{K2} and TESS light curves are in excellent agreement. That was expected since in both light curves the two spots cover very similar latitudes. Compared to the same latitudes on the Sun the $\Delta \Omega'_{\mathrm{Sun}}$ accounts for $\delta \Omega'_{\mathrm{Sun}}\sim14.4\%$ and $\delta \Omega'_{\mathrm{Sun}}\sim13.8\%$ of the full $\Delta \Omega_{\mathrm{Sun},100\%}$ for the \textit{K2} and TESS data, respectively. With these percentages we scaled the rotational shear to the full equator to pole difference, $\Delta \Omega_{*,\mathrm{100\%}}$. All results of the calculations are summarized in Table~\ref{Differential_Rotation}. The weighted average of the \textit{K2} and TESS data resulted in a final value of <$\Delta \Omega$>$=0.118 \pm 0.023$\,rad/d. Our <$\Delta \Omega$> is $\sim$4 times larger than the lower limit in \citet{2016A&A...591A..43D} while the individual values of $\Delta \Omega'$ are $\sim$2 times smaller. If we consider that the long observation baseline of the ASAS data used by \citet{2016A&A...591A..43D} represents a wider range of spot latitudes both results are consistent with each other. Furthermore, our result is in agreement with the findings of \citet{2013A&A...560A...4R} who found horizontal shears up to $\Delta \Omega=0.2$\,rad/d for stars with an effective temperature of $\sim3800$\,K (see their Fig.~15).

\subsubsection{Flares}
\label{EPIC245919787_flares}

As described in Sect.~\ref{flares_EPIC202059229}  we used our earlier published values for the flare frequency and energy of the \textit{K2} data. The flares in the TESS light curve were detected as described in Sect.~\ref{flares}. With our flare finding algorithm we validated 89 and 22 flares for \textit{K2} and TESS, respectively. Fig.~\ref{FFD_EPIC245919787} shows the FFD measured with both instruments. We estimated the completeness limits and found them to be log$(E_{\mathrm{min,K2}})[\mathrm{erg}]=32.24$ and log$(E_{\mathrm{min,TESS}})[\mathrm{erg}]=32.48$. Only flares with energies above this completeness limit were considered in the power-law fit. The best-fitting models are given as the black solid (\textit{K2}) and black dashed (TESS) lines in Fig.~\ref{FFD_EPIC245919787}. The slopes determined for \textit{K2} and TESS data $\beta_{\mathrm{K2}}=-0.84\pm0.06$, $\beta_{\mathrm{TESS,all}}=-0.59\pm0.16$) are significantly different which would imply an enhanced flare frequency in the energy range from log$(E)[\mathrm{erg}]=33-33.5$ in the TESS observations. \citet{2020AJ....159...60G} also analyzed the TESS FFD of EPIC\,245919787 and found a steeper slope that is consistent with our \textit{K2} result. In contrast to the flare detection algorithm of \citet{2020AJ....159...60G} our code can not deal with complex (multi-peak) flares. In fact, by visual inspection it turned out that most of the seven flares above log$(E)[\mathrm{erg}]=33$ show complex flare shapes which results in an overestimation of the flare energy and an underestimation of the flare frequency. We repeated the fitting of the TESS FFD using only flares below log$(E)[\mathrm{erg}]=33$ and above the completeness limit. The resulting power-law slope, $(\beta_{\mathrm{TESS,logE<33}}=-0.83\pm 0.62)$ (shown as the gray dash-dotted line in Fig.~\ref{FFD_EPIC245919787}), is now in excellent agreement with that found for the \textit{K2} data and with the TESS results of \citet{2020AJ....159...60G}. Note, that the three highest energy flares in the \textit{K2} light curve show also complex shapes. However, the error bars for these flares are so large that these points do not affect the power-law-fit.

The specific flare rates we computed for EPIC\,245919787 from the \textit{K2} and TESS light curves are $\nu_{32.5,\mathrm{K2}}=0.46\,\mathrm{d}^{-1}$ and $\nu_{32.5,\mathrm{TESS}}=0.60\,\mathrm{d}^{-1}$. The flare rate of the TESS observations is slightly higher as expected from the higher variability amplitude that points to a higher level of stellar activity in the TESS data.

\begin{figure}
  \centering
  \includegraphics[width=0.35\textwidth,angle=270]{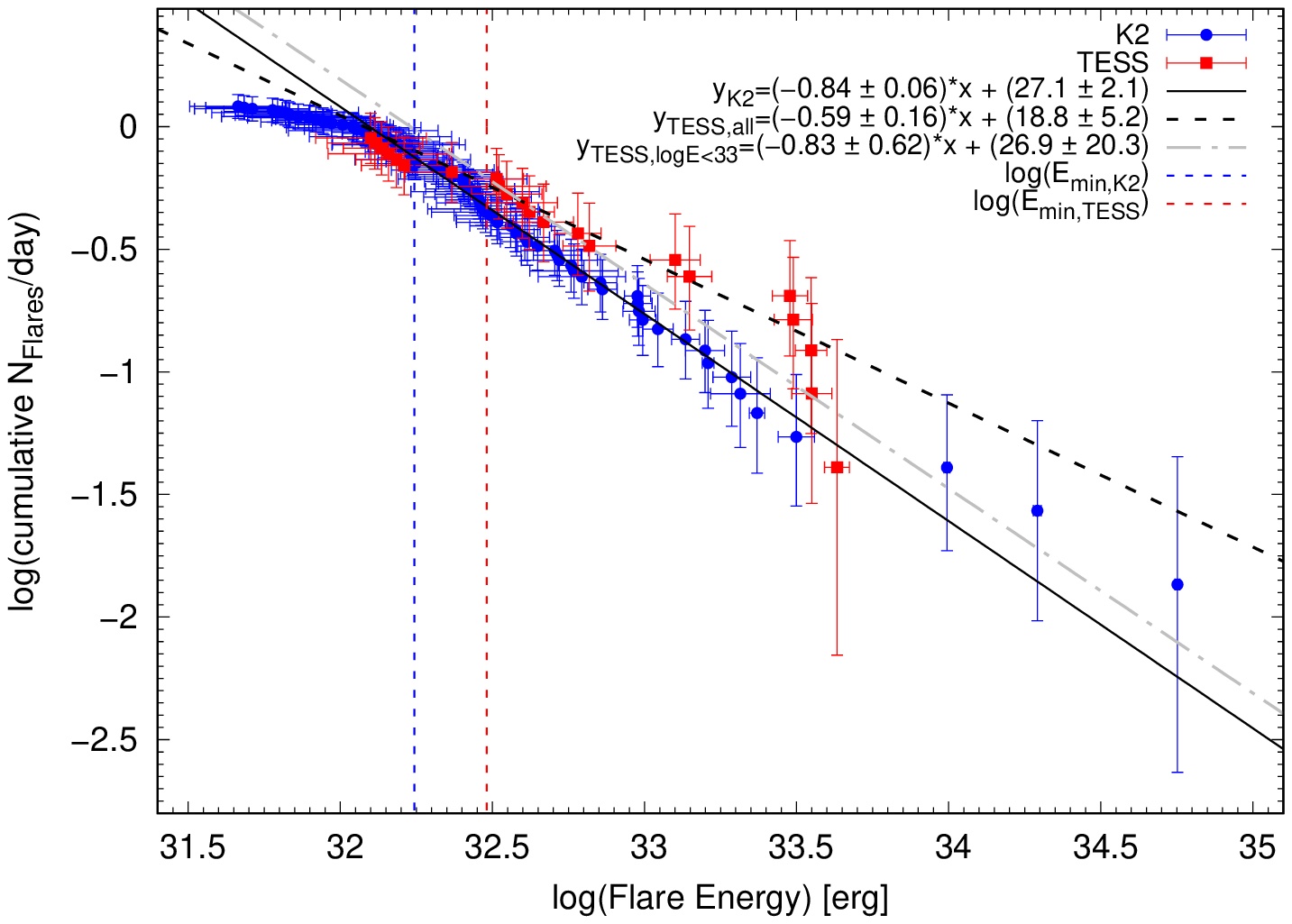}
  \caption{Cumulative flare frequency distribution for the \textit{K2} and TESS data of EPIC\,245919787. The completeness limits are marked as blue and red dashed lines. The best-fitting power-law is given as the black solid (\textit{K2}) and black dashed (TESS) lines. The gray dash-dotted line denotes the fit of the TESS data but excluding complex flares above log$E[\mathrm{erg}]=33$ (see Sect.~\ref{EPIC245919787_flares}). }
  \label{FFD_EPIC245919787}
\end{figure}

%

\section{Summary}
\label{Conclusions}

The photometric space missions \textit{K2} (operating from 2014 - 2018) and TESS  (since 2018) provide a unique opportunity for a precision study of long term stellar variability. Despite they focused on largely exclusive portions of the sky, there is a small region close to the ecliptic that were observed by both, the \textit{K2} and the TESS prime mission (2018-2020). This marginal overlap includes 45 stars from our sample of bright and nearby M dwarfs from the catalog of \citet{2011AJ....142..138L} that were observed by \textit{K2} of which two early M type stars were analyzed in more detail.

We have presented a detailed study on the the long term evolution of rotation periods, star spot distributions and several indicators of stellar activity like flare rates and spot induced variability amplitudes for the two M dwarfs, EPIC\,202059229 and EPIC\,245919787. By using similar data sets from the \textit{K2} and TESS prime missions we are able to analyze the magnetic activity on almost half a decade timescale. This long observational baseline allows us to provide constraints on putative activity cycles and, due to the different location of star spots in the two data sets, also on differential rotation of our target stars.

EPIC\,202059229 is a M1.0V field dwarf that is known to show rotational variability from ASAS observations. On the contrary is the M1.0V dwarf EPIC\,245919787 a member of the $\beta$ Pic moving group and, hence, has a young age of $\sim$24\,Myrs. For both stars the determined rotation period did not change significantly in the $\sim$4.5 and $\sim$2 years elapsed between the \textit{K2} and the TESS observation for EPIC\,202059229 and EPIC\,245919787, respectively. However, for EPIC\,245919787 the rotational cycle shifted in phase by $\sim$0.2 while it stayed constant for EPIC\,202059229. A different behavior was also observed for the amplitude of the rotational modulation. The average variability amplitude changed by $\sim5\sigma$ and $\sim13\sigma$ for EPIC\,202059229 and EPIC\,245919787, respectively. This more significant change in variability amplitude of the $\beta$ Pic member indicates stronger variations of the spot filling factor on the stellar surface which we confirmed by our spot simulations. For both stars we could identify an evolution in the amplitudes for a given light curve that could hint to activity cycles in these stars. The short durations of the individual light curves and the analysis of only two `snapshots' of a putative cycle are not sufficient to detect a periodicity. In the case of EPIC\,245919787 the data are, however, consistent with the cycle period reported by \citet{2017A&A...606A..58D}.

The shape of the phase-folded light curves of both stars provide evidence for star spot evolution within a given observation. From the spot modeling we found that the two assumed spots are located on average between $65^\circ$ and $80^\circ$. With the assumption of the solar rotational profile we were able to determine the differential rotation as a rotational shear. While we could not measure differential rotation for EPIC\,202059229 we found a value for EPIC\,245919787 that is significant on a $\sim10\sigma$-level. This result is consistent with previous studies of EPIC\,245919787 \citep{2016A&A...591A..43D} and of M dwarf with similar effective temperatures \citep{2013A&A...560A...4R}.

Both stars have a rotation period of $\sim5-6$\,d and, hence, are in the saturated regime of the rotation-activity relation. The X-ray luminosity in the 0.1–2.4 keV band \citep[][converted from their X-ray flux using the distance from Table~\ref{stellar_param}]{2022A&A...664A.105F} of EPIC\,245919787 (log$L_{x}[\mathrm{erg/s}]=29.66$) is somewhat higher than the value for EPIC\,202059229 (log$L_{x}[\mathrm{erg/s}]=29.41$), but both are well within the spread of $L_{\rm x}$ in the saturated regime \citep{2020A&A...638A..20M}. \citet{2020A&A...638A..20M} showed that for early-M stars the decay of X-ray activity sets in at $500$\,Myr. Hence, also EPIC\,202059229 seems to be young with an age well below 1\,Gyr. The two stars that have a similar mass are at interesting evolutionary stages, the late pre-main-sequence in the case of EPIC\,245919787 and the young main-sequence in the case of EPIC\,202059229. Differences in the characteristics of their activity might, therefore, be attributed to an evolutionary effect. 

In total, we found 64 and 111 flares for EPIC\,202059229 and EPIC\,245919787, respectively, from which we constructed a cumulative flare frequency distribution for each individual observation. The slopes of the fitted power-laws of all FFDs are consistent with each other and with the average value for stars with spectral types of K7 to M1 measured by \citet{2020A&A...637A..22R}. The specific flare rate, $\nu_{32.5}$, determined in the \textit{K2} and the TESS light curves of EPIC\,202059229 and EPIC\,245919787 follow the same trend as the average variability amplitude. Therefore, variations in flare rates can also act as a  promising method for detecting stellar activity cycles, since the flare rate for the Sun is observed to vary by roughly an order of magnitude between solar maximum and minimum \citep[e.g.,][]{2002A&A...382.1070V,2012ApJ...754..112A}. The potential of this method was shown by \citet{2019RNAAS...3..137S} using 4 years of \textit{Kepler} observations of the M3 dwarf KIC\,8507979. For our targets the specific flare rate is 2-3 times higher for EPIC\,245919787 than for EPIC\,202059229 which confirms that the younger star tends to show an higher level of stellar activity.

In summary, we found that the 24\,Myr young star EPIC\,245919787 shows a higher level of activity with a more significant change of several activity indicators over time than the somewhat older star EPIC\,202059229. Our result is in agreement with recent studies of declining flare activity with age shown in young open clusters \citep[e.g.,][]{2019A&A...622A.133I} and for flare stars in the \textit{Kepler} field \citep{2019ApJ...871..241D}.

\begin{acknowledgements} 
This paper includes data collected with the Kepler and the TESS mission, obtained from the MAST data archive at the Space Telescope Science Institute (STScI). Funding for the TESS mission is provided by the NASA Explorer Program. Funding for the Kepler mission is provided by the NASA Science Mission Directorate. STScI is operated by the Association of Universities for Research in Astronomy, Inc., under NASA contract NAS 5–26555. This research has made use of the Washington Double Star Catalog maintained at the U.S. Naval Observatory.
\end{acknowledgements}


\bibliographystyle{aa}
\bibliography{literatur}

\appendix

\section{Spot modeling: highly degenerate problem}
\label{degeneracy}

The model of the light curve resulted from a rotating star covered with stellar spots is based of many different parameters that are either connected to the properties of the star or the active regions. In such a constellation a unique solution can only be found if information from several different observations are available. The characteristic of this problem is, that parameters are strongly correlated to each other. A change in one of the parameters can be compensated by changing another parameter accordingly. Figure~\ref{FigDegeneracy} gives an example from \begin{small}SOAP 2.0\end{small} for the degeneracy on the TESS light curve of EPIC\,245919787/TIC\,434101713. A dark spot in the center of the stellar disk with a certain size results in an identical light curve than a dark spot at another latitude of an inclined star. Another similar light curve can be obtained by a dark spot in the center of the stellar disk with a different size. Because we do not have access to additional information other than the photometric data, we solved the degeneracy by making suitable assumptions that allowed us to obtain physical light curve models.

   \begin{figure}
   \centering
   \includegraphics[width=0.3\textwidth,angle=270]{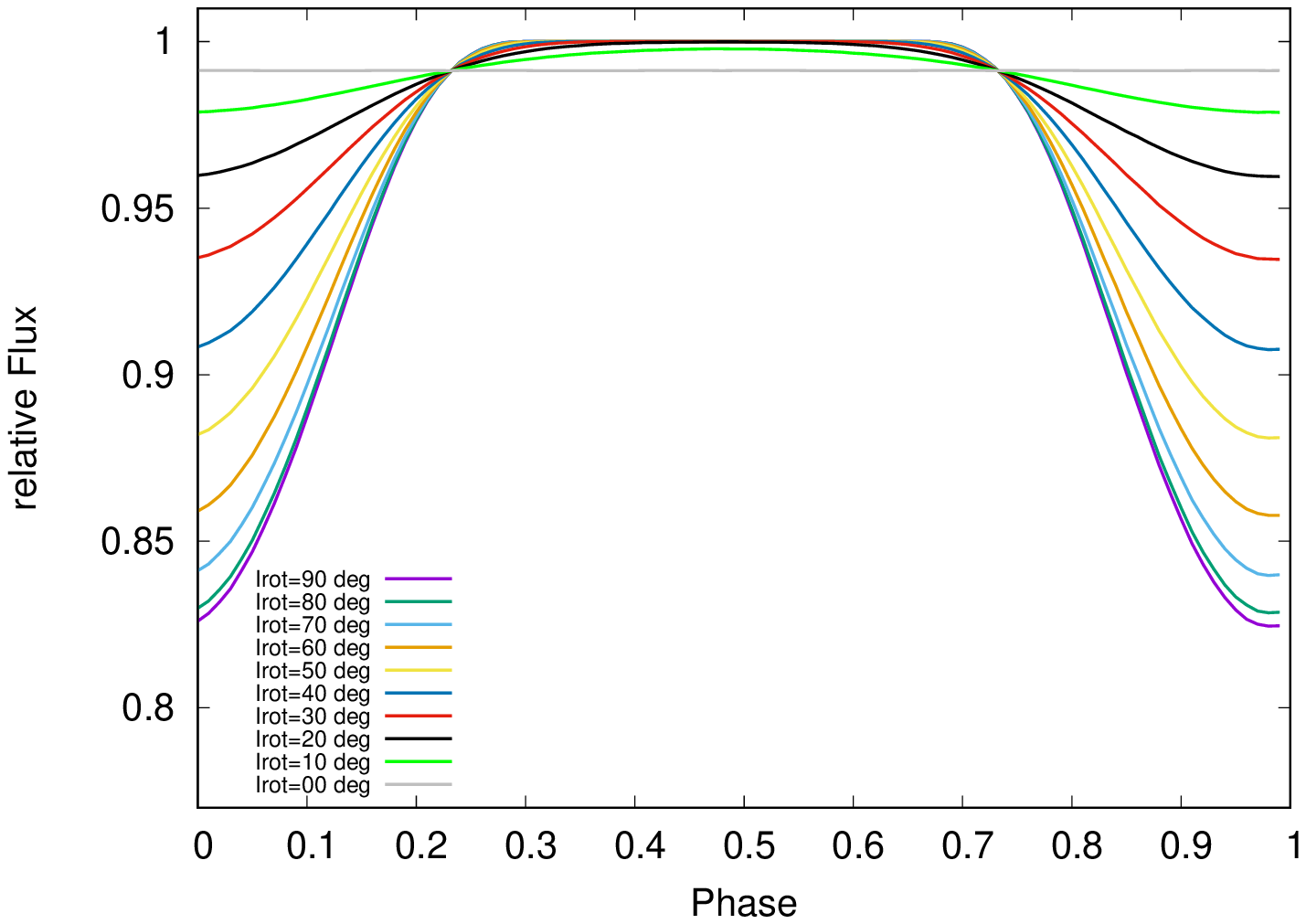}
   \includegraphics[width=0.3\textwidth,angle=270]{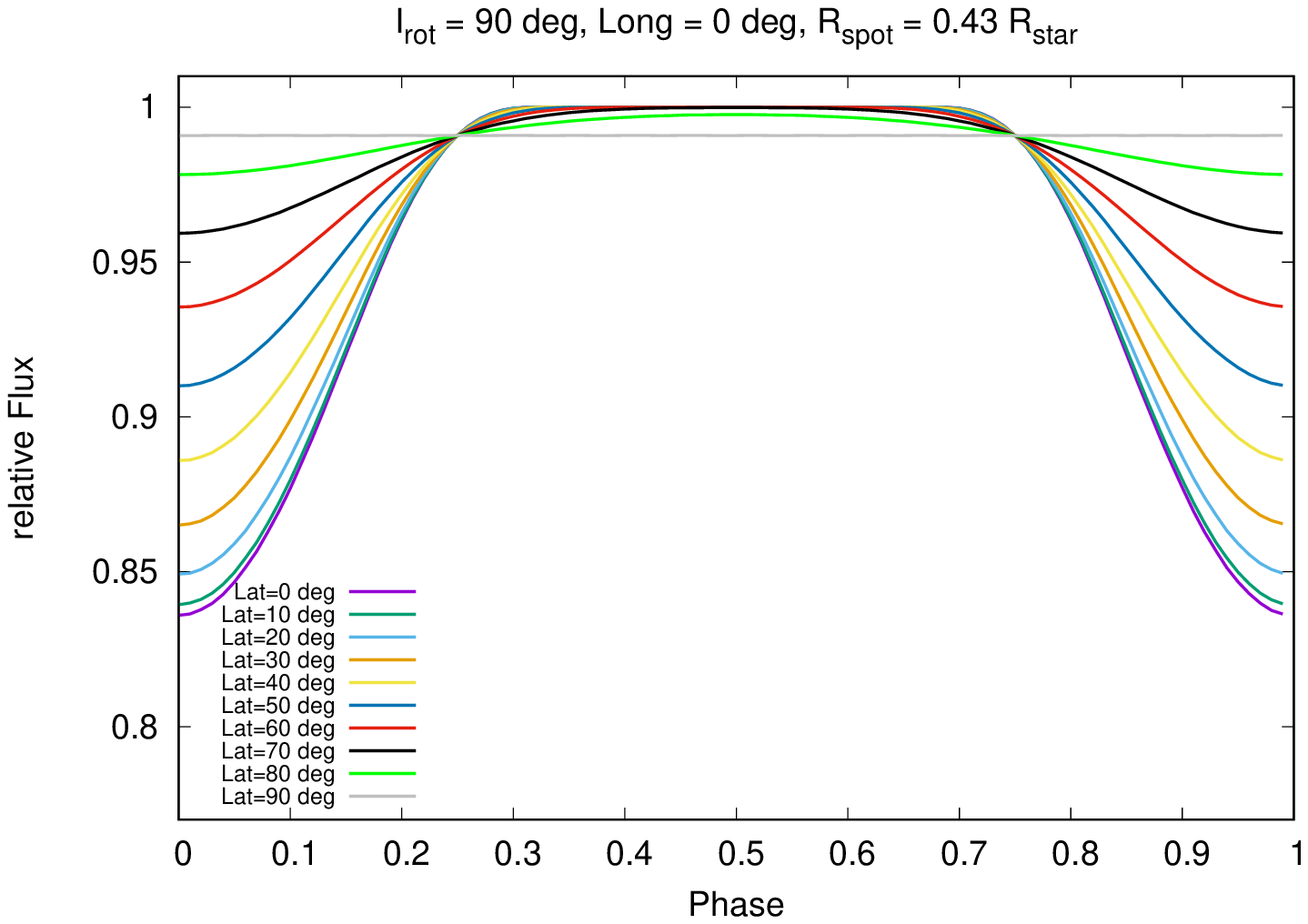}
   \includegraphics[width=0.3\textwidth,angle=270]{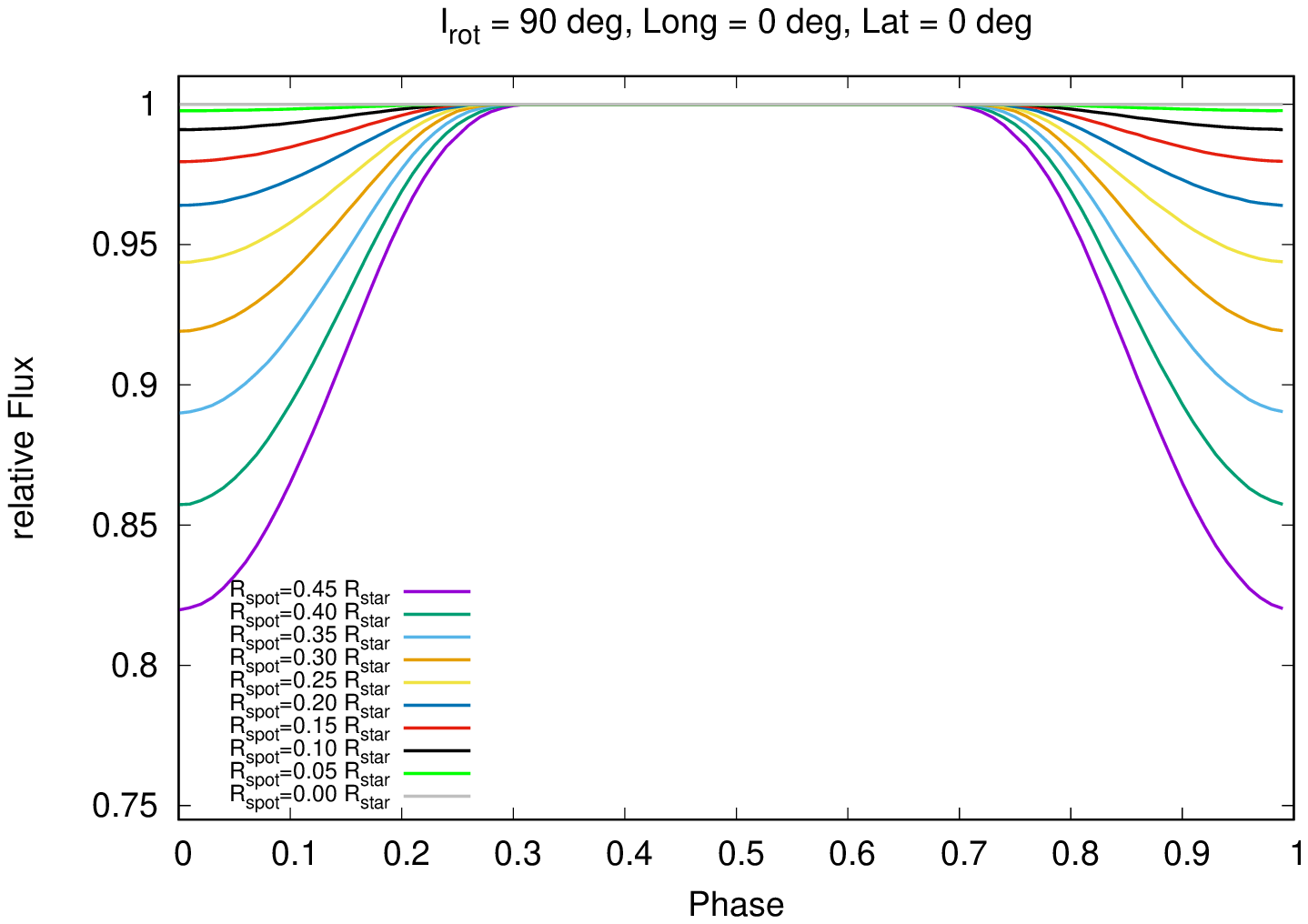}
      \caption{Visualization of the parameter degeneracy on the example of the TESS light curve of EPIC\,245919787/TIC\,434101713. \textbf{Top:} Models for one dark spot with a size of $\frac{R_{\mathrm{spot}}}{R_{\mathrm{star}}}=0.43$ located at longitude=0$^{\circ}$ and latitude=0$^{\circ}$ for different values of the stellar inclination. \textbf{Middle:} Models for one dark spot with a size of $\frac{R_{\mathrm{spot}}}{R_{\mathrm{star}}}=0.43$ located at longitude=0$^{\circ}$ and stellar inclination=90$^{\circ}$ for different values of the latitude. \textbf{Bottom:} Models for one dark spot with a located at longitude=0$^{\circ}$ and latitude=0$^{\circ}$ with a stellar inclination=90$^{\circ}$ for different values of the spot size.}
         \label{FigDegeneracy}
   \end{figure}

\section{Period search results}
\label{period_plots}

For our initial period search we used three different methods, a generalized Lomb-Scargle periodogram, the autocorrelation function (ACF), and fitting the light curves with a sine function as explained in Sect.~\ref{period}. The result of our period search is summarized in plots that contain six panels with the the phase-folded light curve for each method, the periodogram, the ACF and the original light curve with the sine fit. For EPIC\,202059229 Fig.~\ref{Phasfolded_LC_0202059229_K2} shows the result for the \textit{K2} data while in Fig.~\ref{Phasfolded_LC_0202059229_TESS} the results for TESS are given. Fig.~\ref{Phasfolded_LC_0245919787_K2} and Fig.~\ref{Phasfolded_LC_0245919787_TESS} shows the period search results for the \textit{K2} and TESS data of EPIC\,245919787, respectively.

\begin{figure*}
  \centering
  \includegraphics[width=0.6\textwidth,angle=90]{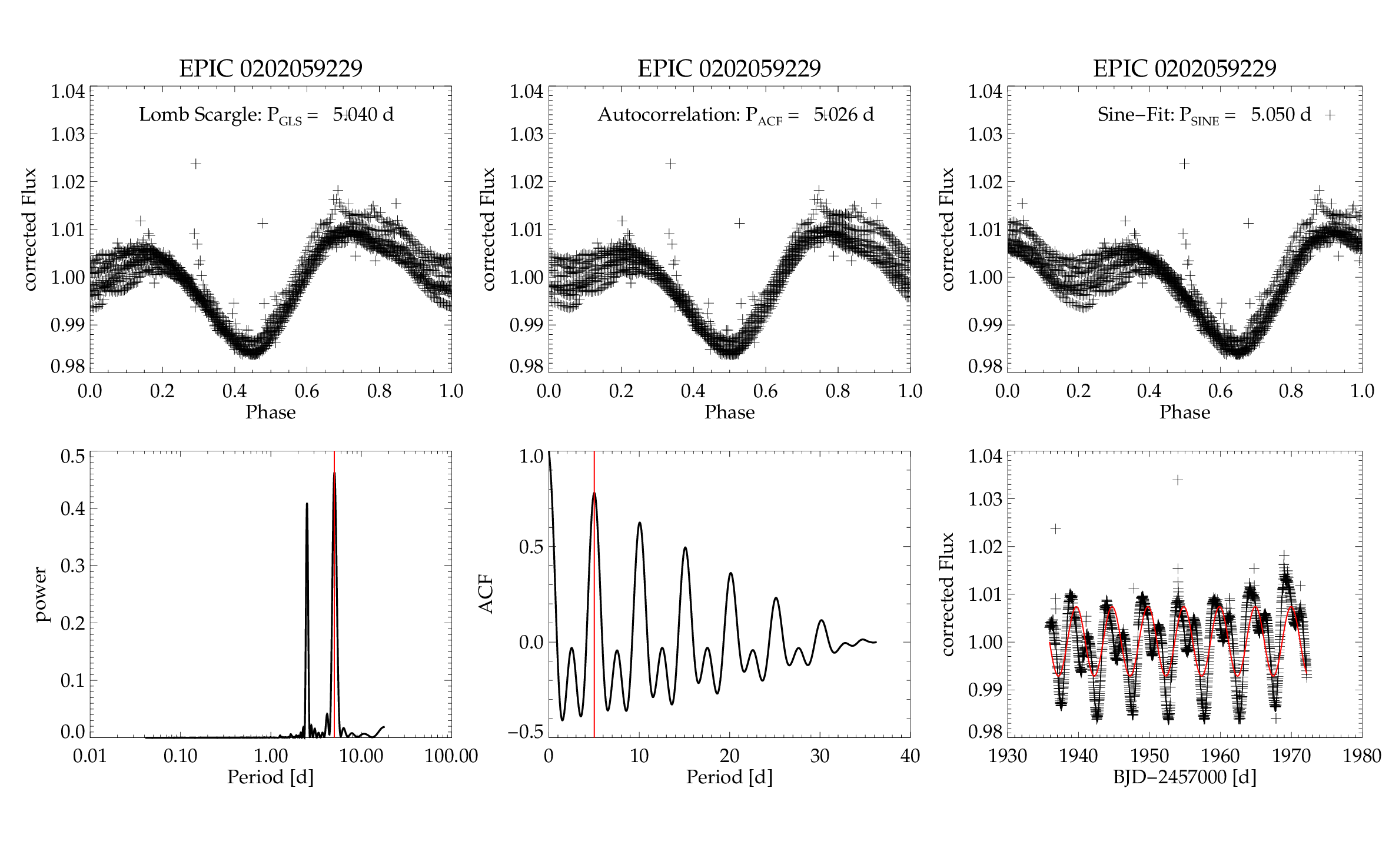}
  \caption{Result of the initial period search for EPIC\,202059229 observed with \textit{K2} in campaign C0. The top panels show the light curves phase-folded with the periods obtained with the different methods. The bottom panel shows the periodogram, the ACF and the original light curve with the sine fit.}
  \label{Phasfolded_LC_0202059229_K2}
\end{figure*}

\begin{figure*}
  \centering
  \includegraphics[width=0.6\textwidth,angle=90]{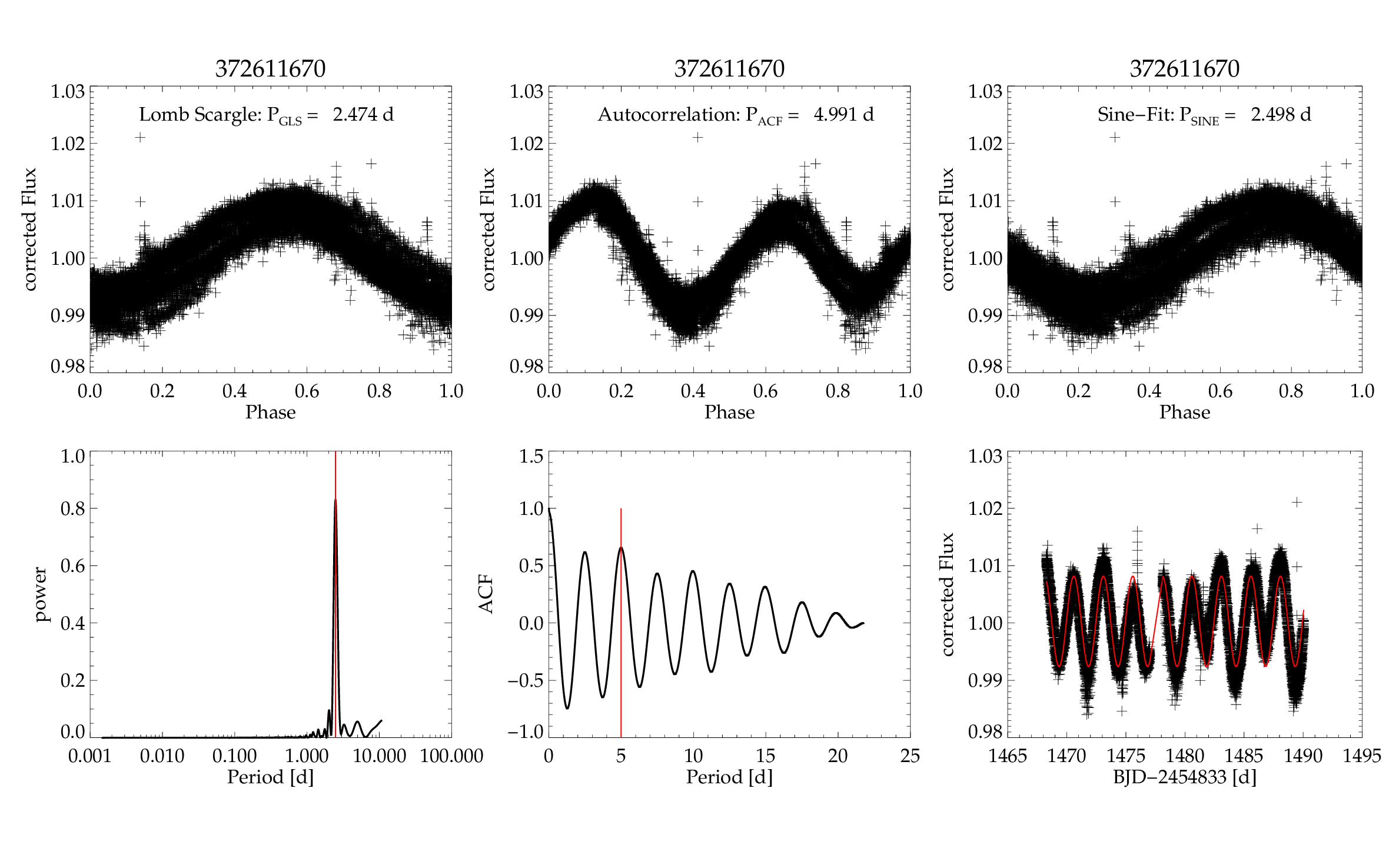}
  \caption{Same as Fig.~\ref{Phasfolded_LC_0202059229_K2} but for the observations of TESS in sector 6. The correct double-humped period for this light curve could only be identified with ACF.}
  \label{Phasfolded_LC_0202059229_TESS}
\end{figure*}

\begin{figure*}
  \centering
  \includegraphics[width=0.6\textwidth,angle=90]{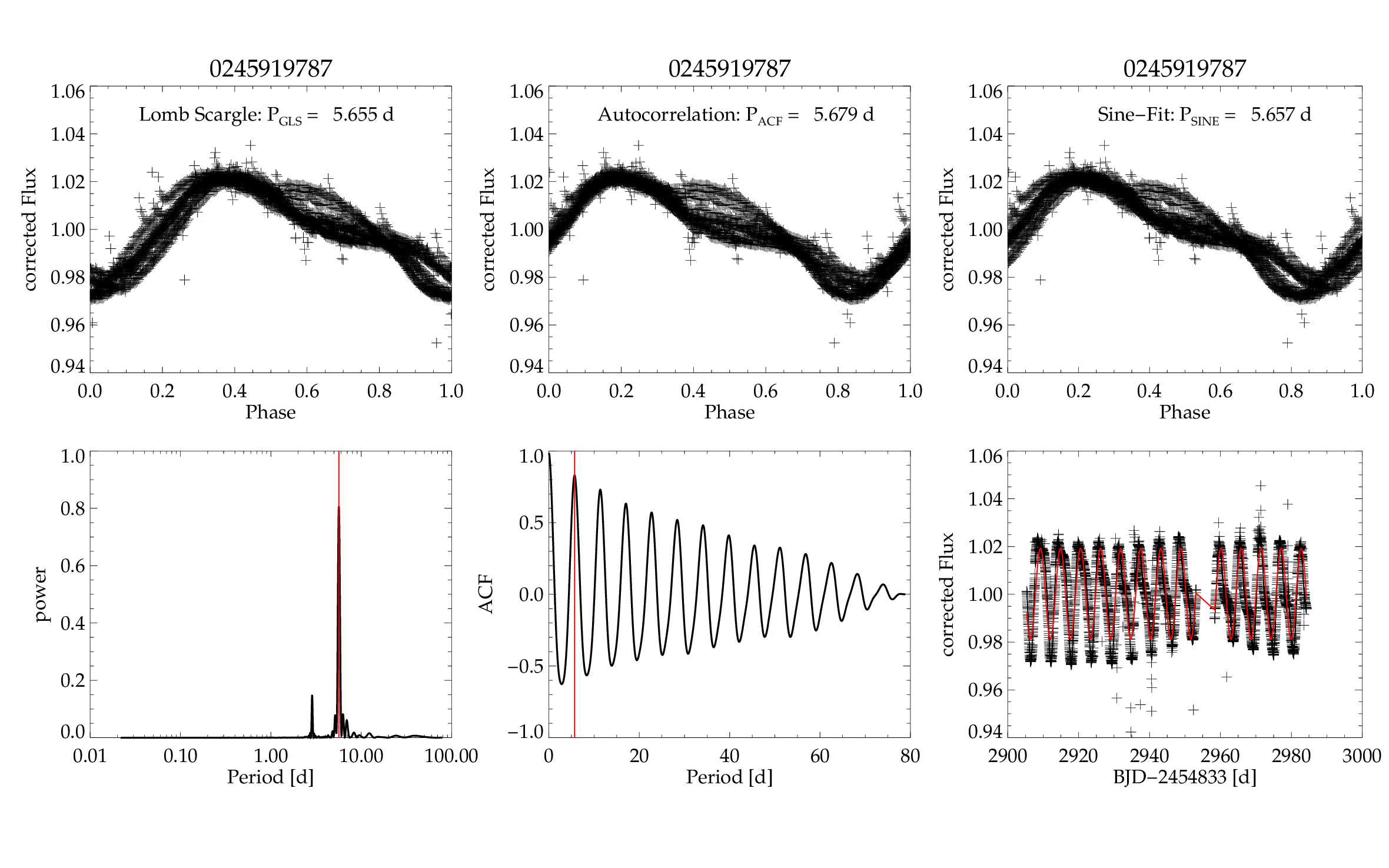}
  \caption{Same as Fig.~\ref{Phasfolded_LC_0202059229_K2} but for or EPIC\,245919787 observed with \textit{K2} in campaign C12.}
  \label{Phasfolded_LC_0245919787_K2}
\end{figure*}

\begin{figure*}
  \centering
  \includegraphics[width=0.6\textwidth,angle=90]{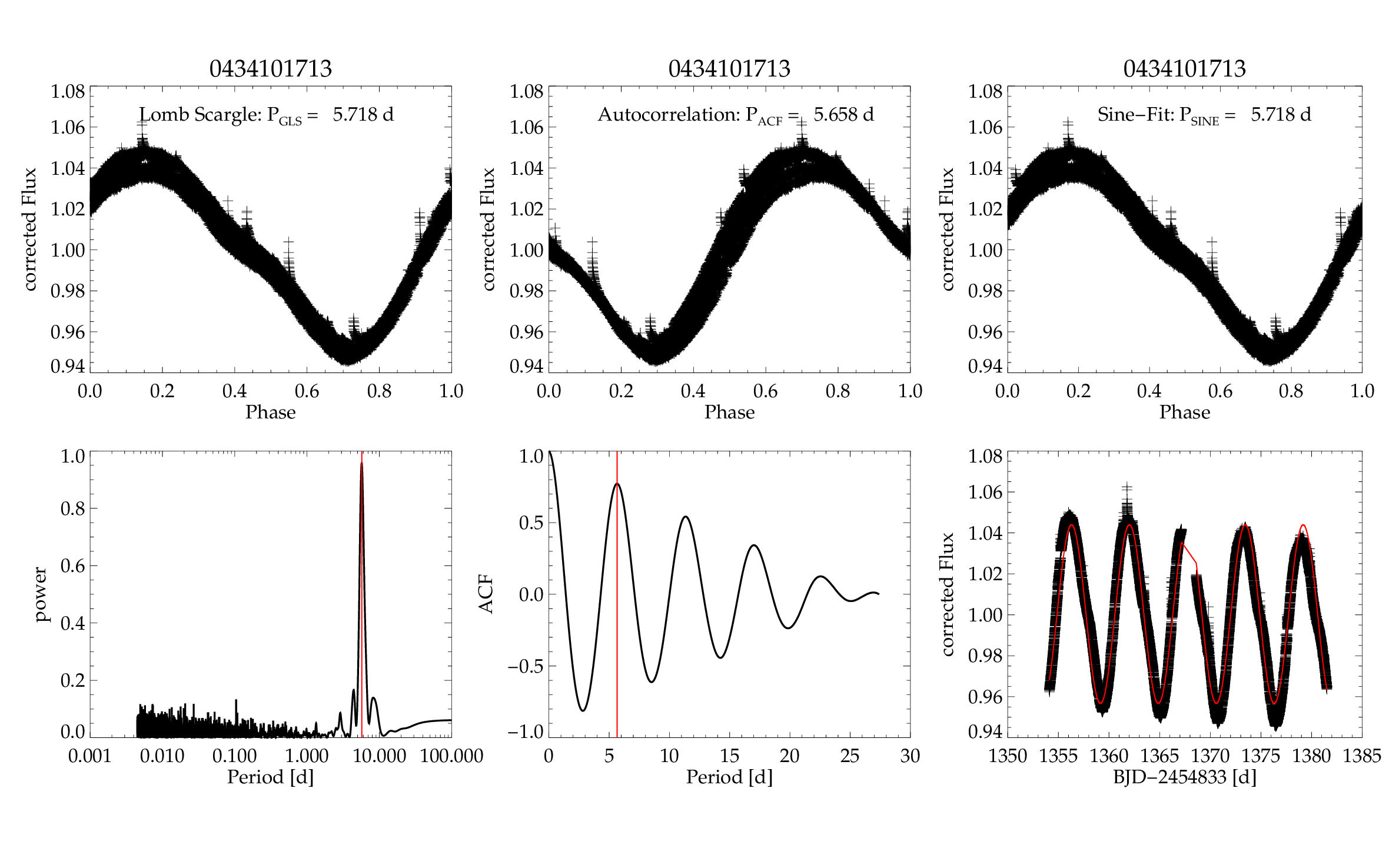}
  \caption{Same as Fig.~\ref{Phasfolded_LC_0245919787_K2} but for the observations of TESS in sector 2.}
  \label{Phasfolded_LC_0245919787_TESS}
\end{figure*}

\end{document}